\documentclass[epsfig,12pt]{article}
\usepackage{amssymb}
\usepackage{amsfonts}
\usepackage{graphicx}
\usepackage{amsmath}

\input epsf.sty
\newtheorem{theorem}{Theorem}
\newtheorem{acknowledgement}[theorem]{Acknowledgement}

\input{tcilatex}
\makeatletter \@addtoreset{equation}{section}
\renewcommand{\theequation}{\thesection.\arabic{equation}}

\begin{document}

\title{%
\rightline{\mbox {\normalsize
{Lab/UFR-HEP0512/GNPHE/0514/VACBT/0514}}} \textbf{Topological }$\mathbf{SL}%
\left( 2\right) $\textbf{\ Gauge Theory on Conifold}}
\author{El Hassan Saidi\thanks{%
h-saidi@fsr.ac.ma} \\
{\small \textit{1.}} {\small \textit{Lab/UFR-Physique des Hautes Energies,
Facult\'{e} des Sciences de Rabat, Morocco.}}\\
{\small \textit{2. Groupement National de Physique des Hautes Energies,
GNPHE; }}\\
{\small \textit{Siege focal, Lab/UFR-HEP, Rabat, Morocco.}}\\
{\small \textit{3. VACBT, Virtual African Centre for Basic Science and
Technology, }}\\
{\small \textit{Focal point Lab/UFR-PHE, Fac Sciences, Rabat, Morocco.}}}
\maketitle

\begin{abstract}
Using a two component $SL\left( 2\right) $ isospinor formalism, we study the
explicit link between conifold $T^{\ast }\mathbb{S}^{3}$ and q-deformed non
commutative holomorphic geometry in complex four dimensions. Then, thinking
about conifold as a projective complex three dimension hypersurface embedded
in non compact $WP^{5}\left( 1,-1,1,-1,1,-1\right) $ space and using
conifold local isometries, we study topological $SL\left( 2\right) $ gauge
theory on $T^{\ast }\mathbb{S}^{3}$ and its reductions to lower dimension
sub-manifolds $T^{\ast }\mathbb{S}^{2}$, $T^{\ast }\mathbb{S}^{1}$ and their
real slices. Projective symmetry is also used to build a supersymmetric QFT$%
_{4}$ realization of these backgrounds. Extensions for higher dimensions
with conifold like properties are explored. \bigskip

\end{abstract}

\tableofcontents

\newpage

\newpage

\section{Introduction}

\qquad In the last decade there has been an intensive interest to
supersymmetric field theories embedded in $10D$ type II superstring models
on Calabi-Yau manifolds. Studies involving conifold backgrounds have been
shown particularly interesting and are of basic importance. They are behind
the derivation of many results in superstring compactification and brane
physics. It is worthwhile to recall the correspondence between conifold and
two dimensional $c=1$ non critical string with cosmological term $\cite{1}$-$%
\cite{5}$, conifold transitions, branes and fluxes, open/closed string
duality $\cite{6}$-$\cite{10}$; and recent development in topological string
theory and non commutative geometry $\cite{11}$-$\cite{17}$.

\qquad Motivated by similarities with non commutative Chern-Simons gauge
theory on 3-sphere and fractional quantum Hall fluids in higher dimensions,
we consider in this paper conifold local isometries and use $SL\left(
2\right) $ isospinor formalism to study non commutative topological gauge
theory on $T^{\ast }\mathbb{S}^{3}$ and reductions to its sub-manifolds $%
T^{\ast }\mathbb{S}^{2}$ and $T^{\ast }\mathbb{S}^{1}$ as well as their
compact real slices. To that purpose, we first explore the relation between
conifold and non commutative geometry. This link, which is already visible
at the level of conifold defining equation $x_{1}y_{2}-x_{2}y_{1}=\mu $, may
be used to develop new perspectives in non commutative gauge theory. A
typical example in this matter is given by the D string fluid model studied
in $\cite{ads}$ and which generalizes FQH systems in Laughlin state with
filling fraction $\frac{1}{k}$. Then, thinking about conifold as a
projective complex three dimension hypersurface embedded in non compact $%
WP^{5}\left( {\small 1,-1,1,-1,1,-1}\right) $, we develop an explicit method
to derive non commutative holomorphic topological gauge theory on conifold
with local $SL\left( 2\right) $ isometry as gauge group. This topological
field theory has the remarkable property of extending Chern-Simons gauge
theory on the 3-sphere. But before describing the organisation of this study
and go into technical details, it is interesting to give other motivations
behind this study. These are given by the three following:

(\textbf{1}) For the link between conifold and non commutative geometry, one
may think about conifold defining equation, 
\begin{equation}
x_{1}y_{2}-x_{2}y_{1}=\mu ,
\end{equation}%
with complex modulus $\mu $, as a typical q-deformed relation of non
commutative geometry (NC) $\cite{18}$-$\cite{23}$. This equation
supplemented by the obvious ones $\left[ x_{i},x_{j}\right] =0$ and $\left[
y_{i},y_{j}\right] =0$, which read altogether in $SL\left( 2\right) $
covariant form as $\varepsilon ^{ij}x_{i}y_{j}=\mu $, $\varepsilon
^{ij}x_{i}x_{j}=0$, $\varepsilon ^{ij}y_{i}y_{j}=0,$ can be equally put in
the form, 
\begin{equation}
x_{[i}y_{j]}=\vartheta _{ij},\qquad x_{[i}x_{j]}=0,\qquad
y_{[i}y_{j]}=0,\qquad i,j=1,2,
\end{equation}%
with a constant \textquotedblleft\ complex magnetic\
field\textquotedblright\ $\vartheta _{ij}\sim \mu \varepsilon _{ij}$ and
where $\left[ ij\right] $ refers to usual antisymmetrisation of indices.
Setting $x_{i}=Z_{1i}$ and $y_{j}=Z_{2j}$, the above relations combine as $%
Z_{ki}Z_{lj}-Z_{kj}Z_{li}=\varepsilon _{kl}\vartheta _{ij}$ and may be read
also as,%
\begin{equation}
Z_{ki}Z_{lj}-\mathcal{R}_{kl}^{mn}Z_{mj}Z_{ni}=\varepsilon _{kl}\vartheta
_{ij},
\end{equation}%
with $\mathcal{R}_{kl}^{mn}=\varepsilon _{k}^{m}\varepsilon _{l}^{n}$.
Equation $x_{1}y_{2}-x_{2}y_{1}=\mu $ is then just the unique non trivial
relation of complex $2\times 2$ matrix coordinate system. With such a
formulation, one disposes of an other picture of thinking about conifold;
and so one can borrow techniques and results on q-deformed non commutative
geometry and symplectic manifolds to build new representations for conifold
and its sub-manifolds. This view offers as well a new way to look for
geometric extensions with conifold like features type the symplectic
varieties with $SP\left( n\right) $ isometries and Nambu like geometry
considered in discussion section.

(\textbf{2}) Conifold geometry seen as a projective hypersurface may be used
to construct supersymmetric QFT$_{4}$ realizations embedded in type II
superstring on conifold. Recall that from super QFT$_{4}$ view, the
projective gauge invariance is the abelian gauge sub-symmetry in
supersymmetric quiver gauge theories. However by describing conifold using
the equation $x_{1}y_{2}-x_{2}y_{1}=\mu $, one is in fact thinking about it
as a complex three dimension holomorphic hypersurface embedded in complex
four dimension space $\mathbb{C}^{4}$ where projective symmetry is fixed. To
implement projective invariance, one needs to go beyond $\mathbb{C}^{4}$;
for instance to four complex dimension non compact projective spaces $W%
\mathbb{P}^{4}\left( {\small -1,1,-1,1,-1}\right) $ where $\mathbb{C}^{4}$
appears as a local patch described by the projective gauge fixing $\sigma =1$%
. The extra variable $\sigma $ captures the projective symmetry of $W\mathbb{%
P}^{4}$ space; that is $\sigma \equiv \lambda \sigma $ with the usual
projective parameter $\lambda \in \mathbb{C}^{\ast }$. Relaxing the
condition $\sigma =1$ to arbitrary values $\sigma \in \mathbb{C}^{\ast }$
and imposing projective invariance, one can easily get the projective
hypersurface describing conifold geometry; but this time embedded in $W%
\mathbb{P}^{4}\left( q_{\sigma },q_{x},q_{y},q_{z},q_{w}\right) $. A quick
way to determine the projective weights $\left( q_{\sigma
},q_{x},q_{y},q_{z},q_{w}\right) $ is to start from eq(\ref{01}) and rewrite
it as 
\begin{equation}
\left( \frac{x_{1}}{\sigma }\right) \left( \sigma y_{2}\right) -\left(
\sigma x_{2}\right) \left( \frac{y_{1}}{\sigma }\right) =\mu .
\end{equation}%
Renaming the variables as $x=\frac{x_{1}}{\sigma }$ and so on, we end with
the hypersurface $xy-zw=\mu $ describing the usual $T^{\ast }\mathbb{P}^{1}$
fibration over $\mathbb{C}^{\ast }$ embedded in the non compact projective\
space $W\mathbb{P}^{4}\left( {\small -1,1,-1,1,-1}\right) $. To keep
Calabi-Yau condition manifest, one should go a step further beyond $W\mathbb{%
P}^{4}\left( {\small -1,1,-1,1,-1}\right) $ and think about above relation
as given by the complex three dimension hypersurface, 
\begin{eqnarray}
\left( \sigma _{+}x_{1}\right) \left( \sigma _{-}y_{2}\right) -\left( \sigma
_{-}x_{2}\right) \left( \sigma _{+}y_{1}\right) &=&\mu ,  \notag \\
\sigma _{+}\sigma _{-} &=&1,  \label{pl}
\end{eqnarray}%
embedded in $W\mathbb{P}^{5}\left( {\small 1,-1,1,-1,1,-1}\right) $. This
formulation is very instructive as it allows to make an idea on the super QFT%
$_{4}$ realization of this background where $x,$ $y,$ $z$ and $w$ \ are
respectively associated with the moduli of fundamental matter superfields $%
X_{+},$ $Y_{-},$ $Z_{+}$ and $W_{-}$. The $\pm $ sub-indices refer to the
projective charge carried by the superfields. Concerning the extra variables 
$\sigma _{\pm }\in \mathbb{C}^{\ast }$, they are associated with two chiral
superfields $\Sigma _{-}$ and $\Sigma _{+}$ constrained as $\Sigma
_{-}\Sigma _{+}=1$.$\ $In this picture, the usual neutral adjoint matter
chiral superfield $\Phi $ of supersymmetric quiver gauge theories appears as
the Lagrange superfield implementing the constraint eq $\Sigma _{-}\Sigma
_{+}=1$ in the holomorphic field action.

(\textbf{3}) The other motivation deals with the relation between conifold
geometry and non commutative topological $SL\left( 2\right) $ gauge theory.
Notice that under local change $x_{i}\rightarrow \Upsilon _{i}^{k}y_{k}$ and 
$y_{j}\rightarrow \left( \Upsilon ^{-1}\right) _{j}^{l}x_{l}$ with $\Upsilon
=\Lambda \left( x_{i},y_{i}\right) $ and $\det \Upsilon =1$, conifold eq $%
\varepsilon ^{ij}x_{i}y_{j}=\mu $ remains invariant. On the moduli space of
the supersymmetric QFT$_{4}$ vacua, these isometries correspond to a non
commutative topological holomorphic $SL\left( 2,\mathbb{C}\right) $ gauge
theory on conifold. As we will see in section 6, the gauge field constraint
eqs are also the field equations of motion of the topological $SL\left( 2,%
\mathbb{C}\right) $ gauge theory. This result is obviously valid in the
projective representation where conifold is thought of as a complex three
hypersurface embedded in $W\mathbb{P}^{5}\left( {\small 1,-1,1,-1,1,-1}%
\right) $. The novelty is that in present case, we have a special abelian
sub-isometry which may be used to get more insight in the huge topological $%
SL\left( 2,\mathbb{C}\right) $ gauge theory and its reductions to the lower
dimension holomorphic gauge theories on $T^{\ast }\mathbb{S}^{2}$ and $%
T^{\ast }\mathbb{S}^{1}$ as well as on real slices. In the abelian sector,
the usual global projective symmetry, $\sigma \rightarrow \lambda \sigma $
and so on, get promoted to a gauge symmetry $\sigma \rightarrow \Lambda
\sigma ,...,$ with $\Lambda =\Lambda \left( \sigma ,x,y\right) $. By
focusing on this abelian gauge sub-symmetry, we show how non commutative
topological $\mathbb{C}^{\ast }$ gauge theory follows from a simple gauge
principle relying on equating the global $SL\left( 2\right) $ algebra, 
\begin{equation}
\left[ D_{+},D_{-}\right] =2D_{0},\qquad \left[ 2D_{0},D_{\pm }\right] =\pm
2D_{\pm },
\end{equation}%
with the corresponding gauge covariant one namely, 
\begin{equation}
\left[ \mathcal{D}_{+},\mathcal{D}_{-}\right] =2\mathcal{D}_{0},\qquad \left[
2\mathcal{D}_{0},\mathcal{D}_{\pm }\right] =\pm 2\mathcal{D}_{\pm }.
\end{equation}%
Here the $\mathcal{D}_{0,\pm }$'s are the covariant derivatives and read as $%
\mathcal{D}_{0,\pm }=D_{0,\pm }-A_{0,\pm }$. Non commutative topological
holomorphic gauge theory on conifold follows naturally as a solution of
these constraint eqs. Chern Simons gauge theory on $\mathbb{S}^{3}$ follows
as well by imposing reality condition.

\qquad Along with these motivations, it is interesting to note moreover that
above mentioned conifold features have similar ones in quantum Hall systems
and attractor mechanism of Hartlee-Hawking wave function for flux
compactifications $\cite{24,25}$. The non commutative topological gauge
theory for conifold may be related to Susskind proposal on quantum Hall (QH)
systems $\cite{26}$. Restricting conifold $T^{\ast }\mathbb{S}^{3}$ to its
real three dimensional slice, the $U\left( 1\right) $ Chern-Simons (CS)
gauge theory on $\mathbb{S}^{3}$ may be, roughly speaking, compared with the 
$\left( 2+1\right) $ non commutative CS gauge theory describing fractional
QH systems. This formal similarity is even more striking since there is also
a correspondence between Susskind model for Laughlin state with filling
fraction $\nu =\frac{1}{k}$, with $k$ positive (odd) integer, and the
attractor mechanism of Hartlee-Hawking universe wave function on $\mathbb{S}%
^{3}$ fixing the global complex deformation parameter of the conifold as $%
\mu =k+i\frac{\phi }{\pi }$ with $\phi $ a real modulus $\cite{27}$.

\qquad The organisation of this paper is as follows: In section 2, we review
aspects of conifold as a hypersurface embedded in the ambiant complex space $%
\mathbb{C}^{4}$, a matter to fix the ideas and convention notations. In
section 3, we show that conifold embedded in $\mathbb{C}^{4}$ may be also
viewed as a q-deformed non commutative complex four dimension holomorphic
geometry with a very special antisymmetric field $\vartheta _{il}$. The same
result is valid for $\mathbb{S}^{3}$ embedded in $\mathbb{R}^{4}$ and the
other sub-manifolds. In section 4, we develop a conifold representation
using a complex three dimension projective hypersurface embedded in $%
WP^{5}\left( {\small 1,-1,1,-1,1,-1}\right) $ and give a super QFT$_{4}$
realization. In section 5, we study conifold diffeomorphism using two kinds
of coordinate frames and in section 6, we consider the derivation of non
commutative topological gauge theory by focusing on the $\mathbb{C}^{\ast }$
model. In section 7, we give a conclusion and make discussions regarding
higher dimension extensions.

\section{Conifold as a $\mathbb{C}^{4}$ hypersurface}

\qquad From the point of view of algebraic geometry, complex three dimension
conifold $T^{\ast }\mathbb{S}^{3}$\ with a global complex deformation
parameter $\mu $ is generally defined as a hypersurface $H_{0}=H_{0}\left(
x_{1},x_{2},y_{1},y_{2}\right) $ embedded in the four complex space $\mathbb{%
C}^{4}$ as, 
\begin{equation}
H_{0}:x_{1}y_{2}-x_{2}y_{1}=\mu ,  \label{1}
\end{equation}%
where $x_{1},x_{2},y_{1},y_{2}$ are complex holomorphic coordinates with the
unique restriction given above. Such a relation, which is singular for $\mu
=0$ and corresponding to a shrinking real three sphere,\ appears as the
topological ground ring of two dimensional non critical $c=1$ string theory $%
\cite{1}$ and has a set of isometries from which one can extract precious
informations. To exhibit explicitly these isometries, it is interesting to
go to the $2\times 2$ matrix coordinates representation by using the
correspondence $\mathbb{C}^{4}\sim Mat\left( 2,\mathbb{C}\right) $ and
redefine conifold hypersurface eq $H_{0}$ as given by the determinant of a
complex holomorphic $2\times 2$ matrix $Z$, i.e 
\begin{equation}
\det Z=\left( x_{1}y_{2}-x_{2}y_{1}\right) =\mu ,
\end{equation}%
with, 
\begin{equation}
Z=\left( 
\begin{array}{cc}
x_{1} & x_{2} \\ 
y_{1} & y_{2}%
\end{array}%
\right) ,\qquad Z.Z^{-1}=I.  \label{2}
\end{equation}%
This complex holomorphic matrix representation, which breaks down for $\mu
=0 $, has a manifest $GL\left( 2,\mathbb{C}\right) \sim \mathbb{C}^{\ast }%
\mathbb{\times }SL\left( 2,\mathbb{C}\right) $ automorphism symmetry acting
through changes generated by the following arbitrary $M$ matrix,%
\begin{equation}
Z\rightarrow MZM^{-1}.  \label{3}
\end{equation}%
Strictly speaking, there are two main options for thinking about this matrix 
$Z$; either as a matrix operator acting on an underlying complex two
dimension space $\mathbb{C}^{2}$, or as pure matrix coordinates $Z_{ij}$
parameterizing $\mathbb{C}^{4}$. If forgetting about small details, the two
options are a priori equivalent and the apparent differences is linked with
the way they are handled. Keeping this in mind, let us focus for the moment
on the isometries of the hypersurface $H_{0}$. Since the $\mathbb{C}^{\ast }$%
\ factor is an abelian symmetry, conifold isometries seems at first sight
given by global $SL\left( 2,\mathbb{C}\right) $ symmetries. However, this $%
SL\left( 2,\mathbb{C}\right) $ isometry is just the global part of a huge
gauge symmetry generated by conifold diffeomorphisms $Diff\left( T^{\ast
}S^{3}\right) $ typically captured by local matrices as shown below, 
\begin{equation}
M_{ij}=M_{ij}\left( x_{1},x_{2},y_{1},y_{2}\right) ,\qquad i,j=1,2.
\end{equation}%
As there is no differential operators $\frac{\partial }{\partial x^{i}}$ and 
$\frac{\partial }{\partial y^{i}}$ in the conifold defining eq $%
x_{1}y_{2}-x_{2}y_{1}=\mu $, it does not matter whether $M$ is a constant
matrix or depending on the local variables $x_{i}$ and $y_{i}$. Leaving
these details for later, note that invariance of the conifold hypersurface $%
H_{0}$ follows from the property of the $\det $ mapping which acts on $Z$ as 
\begin{equation}
\det MZM^{-1}=\det Z,  \label{a1}
\end{equation}%
where $M$ stands for an arbitrary $GL\left( 2,C\right) $ gauge
transformation matrix. Note that for gauge transformations restricted to $K$
sitting in $SL\left( 2,C\right) $, any change of $Z$ type%
\begin{equation}
Z_{ij}\rightarrow Z_{ij}^{\prime }=K_{il}Z_{lj},\qquad K\in SL\left(
2,C\right)  \label{2.7}
\end{equation}%
is also a symmetry of the conifold; thanks to the property $\det \left(
MZ\right) =\left( \det K\right) \left( \det Z\right) $. Note also that with
the change $Z^{\prime }=KZ$ , the previous trivial abelian factor $\mathbb{C}%
^{\ast }\simeq GL\left( 2\right) /SL\left( 2\right) $ eq(\ref{a1}) is no
longer a conifold symmetry of eq(\ref{2.7}); it appears as a
\textquotedblleft scaling transformation" which has much to do with the
scaling symmetry used in $\cite{28}$ to study local complex deformations of
conifold dealing with the building of $\mathbb{S}^{3}$ quantum cosmology.
Recall that these local complex are known to model momenta and winding
corrections in $c=1$ non critical string; for details see $\cite{29}$. Since
the main difference between the transformations $Z^{\prime }=MZM^{-1}$ and $%
Z^{\prime \prime }=KZ$ is this $\mathbb{C}^{\ast }$ abelian scaling factor,
we shall drop it in what follows. Nevertheless it is interesting to note
here that there is a second $\mathbb{C}^{\ast }$ symmetry that we will
encounter below and which turns out to play an important role. It comes from
the factorisation of $SL\left( 2\right) $ as the product of $\mathbb{C}%
^{\ast }$ with the complex holomorphic coset $SL\left( 2\right) /\mathbb{C}%
^{\ast }$; then it should not be confused with $\mathbb{C}^{\ast }\simeq
GL\left( 2\right) /SL\left( 2\right) $ we have just disregarded.

\qquad In the $Z_{ij}$ matrix coordinates formalism, conifold symmetries are
then manifestly exhibited. But this is not all the story; along with this
useful property, one discovers moreover a set of basic features that pass
under heard in the usual $\left( x_{1},x_{2},y_{1},y_{2}\right) $ component
formalism. These features captures essential data on conifold geometry and
have interesting physical interpretations. In what follows, we study three
of the special conifold features that seem to us of basic importance in the
understanding of the structure of the field theoretical models relying on
conifold geometry. These features concern the following points:

\qquad (i) Working explicitly the link between conifold, together with its
sub-manifolds $T^{\ast }\mathbb{P}^{1}$, $\mathbb{S}^{3}$ and $\mathbb{S}%
^{2} $, and q deformed non commutative complex four dimension holomorphic
geometry. As it will be explicated later, these geometries are very special
in the sense that their quadratic algebraic geometry equations look like the
q deformed canonical commutation relations of quantum physics opening then
issues for wider applications. Focusing on conifold equation, it is not
difficult to check that $\det Z=\varepsilon ^{ik}\varepsilon
^{jl}Z_{ij}Z_{kl}=\mu $ is equivalent to the specific q-deformed non
commutative geometry relations, 
\begin{equation}
Z_{ki}Z_{lj}-Z_{kj}Z_{li}=\varepsilon _{kl}\vartheta _{ij},  \label{eqa}
\end{equation}%
where $\mu $\ is as before and where $\varepsilon _{ik}$ is the usual
invariant two dimensional antisymmetric tensor.

\qquad (ii) Use the link $T^{\ast }\mathbb{S}^{3}\sim \mathbb{C}^{\ast
}\times T^{\ast }\mathbb{P}^{1}$ between conifold $T^{\ast }\mathbb{S}^{3}$
and cotangent bundle on the projective space $\mathbb{P}^{1}$ to
re-formulate conifold geometry as a complex three dimension projective
hypersurface in non compact space $W\mathbb{P}^{5}\left( {\small %
1,-1,1,-1,1,-1}\right) $ with complex coordinates $\left( \sigma _{+},\sigma
_{-},x,y,z,w\right) $ and a $\mathbb{C}^{\ast }$ projective symmetry as, 
\begin{equation}
\left( \sigma _{+},\sigma _{-},x,y,z,w\right) \rightarrow \left( \lambda
\sigma _{+},\frac{1}{\lambda }\sigma ,\lambda x,\frac{1}{\lambda }y,\lambda
z,\frac{1}{\lambda }w\right) .  \label{ha}
\end{equation}%
This representation plays a fundamental role in constructing supersymmetric
quiver gauge theories with a $U\left( 1\right) $ sub-symmetry. There, the
complex $\left( \sigma _{+},\sigma _{-},x,y,z,w\right) $ variables describe
superfields moduli minimizing the chiral superpotential, 
\begin{eqnarray}
\int d^{2}\theta \mathcal{W} &=&\int d^{2}\theta \left( \left[ X_{+}\Phi
_{0}Y_{-}-Z_{+}\Phi _{0}W_{-}\right] -\mu \Sigma _{+}\Phi _{0}\Sigma
_{-}\right)  \notag \\
&&+\int d^{2}\theta \Psi _{0}\left( \Sigma _{+}\Sigma _{-}-1\right) \\
&&+\int d^{2}\theta \mathrm{W}\left( \Sigma _{\pm
},X_{+},Y_{-},Z_{+},W_{-}\right) ,  \notag
\end{eqnarray}%
with respect to the neutral chiral matter superfields $\Phi _{0}$ and $\Psi
_{0}$. The sub-indices $\pm $ refer to the gauge charges. The naive
correspondence would be associating the $\sigma _{+},$ $\sigma _{-},$ $x,$ $%
y,$ $z$ and $w$ variables with the VEVs of corresponding superfields $\left(
\Sigma _{-},X_{+},Y_{-},Z_{+},W_{-}\right) $. A priori we may have $%
x=<X_{+}>,$ $y=<Y_{-}>,$ $z=<Z_{+}>,$ $w=<W_{+}>$ and $\sigma _{\pm
}=<\Sigma _{\pm }>$ following from extremizing $\mathrm{W}\left( \Sigma
_{\pm },X_{+},Y_{-},Z_{+},W_{-}\right) $,%
\begin{equation}
d\mathrm{W}\left( \Sigma _{\pm },X_{+},Y_{-},Z_{+},W_{-}\right) =0.
\end{equation}%
However gauge invariance requires that the exact picture should be as
follows 
\begin{eqnarray}
xy &=&<X_{+}Y_{-}>,\qquad zw=<Z_{+}W_{-}>,  \notag \\
\sigma _{-}x &=&<\Sigma _{-}X_{+}>,\qquad \sigma _{-}w=<\Sigma _{-}Z_{+}>, \\
\sigma _{+}y &=&<\Sigma _{+}Y_{-}>,\qquad \sigma _{+}w=<\Sigma _{+}W_{-}> 
\notag
\end{eqnarray}%
More details on this method are given in subsection 4.2. For a quite similar
super QFT$_{4}$ analysis dealing with $T^{\ast }\mathbb{P}^{1}$ background;
see $\cite{30,9}$.

\qquad (iii) Referring to above superfield theoretical interpretation and to
conifold eq $x_{1}y_{2}-x_{2}y_{1}=\mu $,\ which we can usually rewrite it
as follows, 
\begin{equation}
x_{1}y_{2}-x_{2}y_{1}=\left( \frac{x_{1}}{\sigma }\right) \left( \sigma
y_{2}\right) -\left( \frac{x_{2}}{\sigma }\right) \left( \sigma y_{1}\right)
=\mu ,
\end{equation}%
for any non zero complex modulus $\sigma $. This change corresponds to
moving from the $\mathbb{C}^{\ast }$ invariant coordinate frame $\left(
x_{1},x_{2},y_{2},y_{1}\right) $ to the projective one $\left( \sigma
,x,y,z,w\right) $. The objective of this part of the analysis is to extend
the global change (\ref{ha}) to local gauge transformations 
\begin{equation}
\sigma \rightarrow \frac{1}{\Lambda }\sigma ,\qquad \Lambda =\Lambda \left(
\sigma ,x,y,z,w\right) ,
\end{equation}%
which are still isometries of conifold. This local change induces in turns,%
\begin{eqnarray}
x &=&\left( \frac{x_{1}}{\sigma }\right) \rightarrow \Lambda \left( \frac{%
x_{1}}{\sigma }\right) =\Lambda x  \notag \\
z &=&\left( \frac{x_{2}}{\sigma }\right) \rightarrow \Lambda \left( \frac{%
x_{2}}{\sigma }\right) =\Lambda Z  \notag \\
y &=&\left( \sigma y_{1}\right) \rightarrow \frac{1}{\Lambda }\left( \sigma
y_{1}\right) =\frac{1}{\Lambda }y \\
w &=&\left( \sigma y_{2}\right) \rightarrow \frac{1}{\Lambda }\left( \sigma
y_{2}\right) =\frac{1}{\Lambda }w.  \notag
\end{eqnarray}%
Then study the gauge theory behind this gauge invariance principle. As we
will prove in section 5, this is a non commutative holomorphic topological
gauge theory which on real slice, reduces to the non commutative topological
Chern-Simons gauge theory on the three sphere. This topological gauge theory
deals with abelian isometries and is in fact a part of the huge $SL\left(
2\right) $ holomorphic gauge theory.

\section{Conifold as a q-deformed NC $\mathbb{C}^{4}$ geometry}

\qquad In this section, we first introduce the two component formalism to
parameterize conifold geometry. Then, we discuss its link with q-deformed
non commutative geometry in $\mathbb{C}^{4}$. Finally we give a classical
mechanical like model realizing conifold background. This complex
holomorphic model is inspired from similarities with the classical dynamics
of quantum Hall particles moving in a strong magnetic field.

\subsection{Two component formalism}

\qquad As far as conifold defining eq(\ref{1}) is concerned, one learns from
the coordinate matrix representation that if we insist on using component
formalism for $T^{\ast }\mathbb{S}^{3}$, the natural way to do it is by
using a two component formalism involving the two complex holomorphic $%
SL\left( 2,C\right) $ isospinors, 
\begin{equation}
u_{i}=\left( x_{1},x_{2}\right) ,\qquad v_{i}=\left( y_{1},y_{2}\right) .
\label{4}
\end{equation}%
These two component variables are given by the rows of the matrix coordinate 
$Z_{ij}$; that is $u_{i}=Z_{1i}$ and $v_{i}=Z_{2i}$. In terms of these
isospinor variables, conifold constraint eq reads as 
\begin{eqnarray}
\varepsilon ^{ij}u_{i}v_{j} &=&\mu ,  \notag \\
\varepsilon ^{ij}u_{i}u_{j} &=&0, \\
\varepsilon ^{ij}v_{i}v_{i} &=&0,  \notag
\end{eqnarray}%
where $\varepsilon ^{ij}$ is the usual two dimensional antisymmetric
invariant tensor with $\varepsilon ^{12}=1$ and inverse $\frac{1}{2}%
\varepsilon _{ji}$. The first relation $\varepsilon ^{ij}u_{i}v_{j}=\mu $
expresses just $SL\left( 2,C\right) $ invariance of conifold hypersurface in 
$\mathbb{C}^{4}$. The two remaining others rests on the property that $u_{i}$
and $v_{i}$ are commuting bosonic isodoublets in same manner as for Penrose
twistors $\cite{31}$.\ The fact that conifold holomorphic hypersurface $%
H_{0} $ takes the above form seems at first sight something obvious and it
is just a way of exhibiting manifestly $SL\left( 2,C\right) $\ global
isometries. This is true, but there is something more. The idea is that, by
help of the inverse of $\varepsilon ^{ij}$; i.e $\varepsilon
^{ij}\varepsilon _{ji}=2$, these relations may be also put into the
following remarkable form,%
\begin{eqnarray}
u_{i}u_{j}-u_{j}u_{i} &=&0,  \notag \\
u_{i}v_{j}-v_{i}u_{j} &=&\vartheta _{ij},  \label{5} \\
v_{i}v_{j}-v_{j}v_{i} &=&0,  \notag
\end{eqnarray}%
where we have set $\vartheta _{ij}=\frac{\mu }{2}\varepsilon _{ji}$. But
these are familiar relations in non commutative geometry; the only
differences are that in present case we are dealing with complex holomorphic
analysis in higher dimensions and that the precise interpretation is that
the identity $\left( u_{i}v_{j}-v_{i}u_{j}\right) =\vartheta _{ij}$ is q
deformed relation 
\begin{equation}
u_{i}v_{j}-R_{ij}^{kl}v_{k}u_{l}=\vartheta _{ij},
\end{equation}
with $R_{ij}^{kl}=\varepsilon _{i}^{k}\varepsilon _{j}^{l}$. Forgetting
about this technical detail, the two component isospinor formalism we have
been introducing establishes therefore a direct and manifest link between
conifold hypersurface and q-deformed non commutative holomorphic geometry in
complex four dimensions with magnitude of the deformation tensor given by
the global complex deformation $\mu $.

\begin{theorem}
\qquad Conifold $T^{\ast }\mathbb{S}^{3}$ with complex moduli $\mu $ is
equivalent to a q-deformed non commutative complex four dimension geometry
with $SL\left( 2\right) $ isometry and holomorphic magnetic field $B_{IK}=-%
\frac{\mu }{2}\varepsilon _{ik}\varepsilon _{jl}$, $I=\left( i,j\right) $
and $K=\left( k,l\right) $. In this view, the singular limit $\mu =0$
corresponds to commuting $\mathbb{C}^{4}$. This result is also valid for the
complex two dimension holomorphic sub-manifold $T^{\ast }\mathbb{P}^{1}$
having a $SL\left( 2\right) /\mathbb{C}^{\ast }$\ isometry, the real slice $%
\mathbb{S}^{3}$ with $SU\left( 2\right) $ isometry and the two sphere $%
\mathbb{S}^{2}$ with symmetry $SU\left( 2\right) /U\left( 1\right) $.
\end{theorem}

\qquad With this result at hand, one may be tried to do something with;
starting with the search for bonds with relevant quantities in type II
superstring compactifications on Calabi-Yau manifolds and brane physics. A
way to make contact with the real quantities is to restrict conifold $%
T^{\ast }\mathbb{S}^{3}$ to its real slice obtained by setting $y_{2}=%
\overline{x_{1}}$ and $y_{1}=\overline{x_{2}}$. The defining equation of the
resulting real three sphere $\mathbb{S}^{3}$ embedded in $\mathbb{C}^{2}\sim 
\mathbb{R}^{4}$ reads then as,%
\begin{equation}
\left\vert x_{1}\right\vert ^{2}+\left\vert x_{2}\right\vert ^{2}=\func{Re}%
\mu =p,
\end{equation}%
where now the real number $p$ is the radius squared of $\mathbb{S}^{3}$. In
this real restriction, the isospinor $v_{i}$ get identified with $\overline{u%
}_{i}=\left( \overline{u^{i}}\right) $, the complex conjugate of $%
u^{i}=\varepsilon ^{ij}u_{j}$ and eqs(\ref{5}) reduce to the following
special non commutative geometry relations,%
\begin{eqnarray}
u_{i}u_{j}-u_{j}u_{i} &=&0,  \notag \\
u_{i}\overline{u}_{j}-\overline{u}_{i}u_{j} &=&p\varepsilon _{ij}, \\
\overline{u}_{i}\overline{u}_{j}-\overline{u}_{j}\overline{u}_{i} &=&0. 
\notag
\end{eqnarray}%
If we forget about reality and continue to work with complex holomorphic
quantities, one may be tempted to derive new representations for conifold
geometry by mimicking standard analysis in quantum mechanics and non
commutative geometry developed in literature $\cite{32}$-$\cite{33}$. The
global complex deformation parameter $\mu $ has formally a similar role as
Planck constant $\hbar $ of quantum mechanics and so may be used for
quasi-classical studies using formal series in powers of $\mu $ as,%
\begin{equation}
\mathcal{F}\left( Z_{ij}\right) =\sum_{n}\mu ^{n}F_{n}\left( Z_{ij}\right) ,
\end{equation}%
where the $F_{n}\left( Z_{ij}\right) $'s refer to the \textit{n-th}
perturbation order of correction terms. This expansion is in agreement with
the structure of the expansion of free energy $\mathcal{F}_{top}$ of the B
model topological string on locally conifold with local complex deformations.

\subsection{A classical mechanical like model}

\qquad Letting the two component variables $u$ and $v$ to have a time
dependence as $u^{i}=U^{i}\left( t\right) $ and $v_{i}=V_{i}\left( t\right) $
with $\partial _{t}U^{i}$ being the "time" derivative, this pair of
isospinors may be also interpreted as canonical variables in a complexified
dynamical Lagrangian description,%
\begin{equation}
V_{i}\sim \frac{\partial \mathcal{L}}{\partial \left( \partial
_{t}U^{i}\right) },  \label{can}
\end{equation}%
where $\mathcal{L}=\mathcal{L}\left( U,\partial _{t}U\right) $ is some
holomorphic Lagrangian field density. The simplest example is given by the
holomorphic field density, 
\begin{equation}
\mathcal{L}\sim \frac{1}{\mu }\varepsilon ^{ij}V_{i}\partial _{t}U_{j}=\frac{%
1}{\mu }V_{i}\partial _{t}U^{i},  \label{cam}
\end{equation}%
which may be thought of in the same lines as the real lagrangian describing
a two dimensional quantum Hall effect particle moving in an external
perpendicular and strong magnetic field $B\sim \frac{1}{\mu }$. Computing
the conjugate momentum $P_{i}=\frac{\partial \mathcal{L}}{\partial \left(
\partial _{t}U^{i}\right) }$ of the field variable $U^{i}$ by using eqs(\ref%
{can}-\ref{cam}), we find 
\begin{equation}
P_{i}=\frac{1}{\mu }V_{i},
\end{equation}%
which up to using the q deformed canonical commutation relation $%
P_{i}U^{i}-U_{i}P^{i}\sim 1$, we get%
\begin{equation}
\left( V_{i}U^{i}-U_{i}V^{i}\right) \sim \mu
\end{equation}%
which is nothing but conifold equation. One may also consider holomorphic
hamiltonian like representations with,%
\begin{equation}
\mathcal{H}\left( U,V\right) =\sum_{i,j=1}^{2}\varepsilon ^{ij}V_{i}\partial
_{t}U_{j}-\mathcal{L},
\end{equation}%
to develop a symplectic like geometry. We believe that this two component
isospinor formalism, which recalls Pauli two component spinor formalism of
QED, may encode deeper informations on conifold. Together with the explicit
non commutative property, this isospinor formalism opens a new insight for
developing other conifold representations which may be linked with recent
developments in topological string theory on conifold. Two of these
representations, under investigation in $\cite{35},$ are given by the field
theoretic description \`{a} la Susskind $\cite{27}$ and the the matrix model
realization \`{a} la Susskind-Polychronakos $\cite{36,37}$. In the effective
field theory approach, $\func{Re}\mu $ is interpreted as given by the
inverse of an external constant magnetic field $B_{ex}$,%
\begin{equation}
\func{Re}\left( \vartheta _{ij}\right) =\frac{k}{B_{ex}}\varepsilon
_{ij},\qquad k=0,1,2,...,  \label{55}
\end{equation}%
and the used method follows Susskind philosophy in dealing with Quantum Hall
Effect (QHE) as a non commutative effective Chern Simons $U\left( 1\right) $
gauge theory. Recall in passing that in Susskind proposal $\cite{28}$ that
non-Commutative Chern-Simons gauge theory on the ($2+1$) real space provides
a natural framework to study the Laughlin state of filling factor $\nu =%
\frac{1}{k}$ with $k$ a positive (odd) integer. Following $\cite{28,39}$,
the non commutativity parameter $\vartheta $ of the co-moving plane
coordinates is related to the filling factor $\nu $ and to the Chern-Simon
effective field coupling constant $\lambda _{CS}$ as 
\begin{equation}
\nu \times \vartheta \times B_{ex}=\nu \times \lambda _{CS}=1.
\end{equation}%
Upon on rescaling eq(\ref{55}) as $\func{Re}\left( \vartheta _{ij}\right)
=k\varepsilon _{ij}$ and completing $\func{Re}\left( \vartheta _{ij}\right) $
by switching on the imaginary part $\func{Im}\vartheta =\frac{\phi }{\pi }$,
we get a remarkable relation%
\begin{equation}
\mu =k+\frac{i}{\pi }\phi ,\qquad k=0,1,2,...,
\end{equation}%
which should be compared with one of two crucial relations derived in $\cite%
{40}$ and concerning attractor mechanism eqs for Hartlee Hawking universe
wave function,%
\begin{equation}
X^{i}=k^{i}+\frac{i}{\pi }\Phi ^{i},
\end{equation}%
where $k^{i}=\func{Re}\left( X^{i}\right) $ are integers.\ In this equation,
the complex numbers 
\begin{equation}
X^{i}=\int_{A_{i}}\Omega _{3}
\end{equation}%
are the usual complex structures given by integral of holomorphic 3-form $%
\Omega _{3}$ over 3-cycles $A_{i}$ of integral cohomology $H_{3}\left(
T^{\ast }\mathbb{S}^{3},\mathbb{Z}\right) $. The way Susskind model for
quantum Hall systems is related to the attractor mechanism of $\cite{25}$ is
still unclear for us; it needs more exploration.

\qquad To conclude this section, we would like to recall some facts. First
note that appearance of non commutative geometry behavior for conifold is
not a strange feature. It is quite well established that non commutativity
lifts singularity $\cite{21}$; and that deformed conifold has a non
commutative geometry interpretation is then obvious. This property has been
explicitly checked for Calabi-Yau orbifold geometries with discrete torsion
and has been also interpreted in terms of fractional branes $\cite{23}$.
Note also that, even though not extensively explicited, non commutative
behaviour of conifold is understood in the study of topological string
theory on conifold; in particular by using geometric transition and mirror
symmetry between $A$ and $B$ models. There, the complex parameter $\mu $\ of 
$B$ model, given by integral of holomorphic 3-form $\Omega _{3}$, $\mu
=\int_{A_{3}}\Omega _{3}$, over a 3-cycle $A_{3}$ of $H_{3}\left( T^{\ast
}S^{3}\right) $, is mirror to complexified Kahler parameter $%
t=\int_{D_{2}}\left( B_{2}+iK_{2}\right) $ of the topological $A$ model
involving magnetic like field.

\section{Conifold as a projective hypersurface}

\qquad In eq(\ref{1}), conifold is viewed as a hypersurface embedded in $%
\mathbb{C}^{4}$ and the variables $x_{1},$ $x_{2},$ $y_{1}$ and $y_{2}$ were
arbitrary complex holomorphic coordinates in $\mathbb{C}$ with the unique
restriction $x_{1}y_{2}-x_{2}y_{1}=\mu $. From the point of view of super QFT%
$_{4}$ functional analysis, the VEV's of the operator product $\mathcal{F}%
_{1}...\mathcal{F}_{n}$, 
\begin{equation}
<\mathcal{F}_{1}...\mathcal{F}_{n}>=\int_{T^{\ast }\mathbb{S}%
^{3}}\dprod\limits_{i=1}^{2}DX_{i}DY_{i}D\Phi _{0}\left( \mathcal{F}_{1}...%
\mathcal{F}_{n}\right) \exp \mathcal{S}\left[ \Phi _{0},X_{i},Y_{j}\right] ,
\end{equation}%
with $\mathcal{F}\left( \Phi _{0},X_{i},Y_{j}\right) $ a generic function
depending on the chiral superfields $\Phi _{0},X_{i}$ and $Y_{j}$, the eq $%
x_{1}y_{2}-x_{2}y_{1}=\mu $ is recovered from the following global
holomorphic superfield action,%
\begin{equation}
\mathcal{S}\left[ \Phi _{0},X_{i},Y_{j}\right] =\int d^{4}xd^{2}\theta
\left( X_{1}\Phi _{0}Y_{2}-X_{2}\Phi _{0}Y_{1}-\mu \Phi _{0}\right) +\int
d^{4}xd^{2}\theta W\left( X_{i},Y_{i}\right)
\end{equation}%
Notice that while the superfields $X_{i}$ and $Y_{j}$ come in pairs, that is
in $SL\left( 2\right) $ doublets, the chiral superfield $\Phi _{0}$ is a
singlet and appears as a Lagrange superfield parameter. This feature shows
that $\mathcal{N}=2$ supersymmetry spectrum (two hypermultiplets) is
partially broken down to $\mathcal{N}=1$ in agreement with known results on
conifold geometries. Notice also that up to now, we have no gauge symmetry
yet; the superfields $\Phi _{0},X_{i}$ and $Y_{j}$ are not charged. In what
follows, we study the gauging of this model by approaching conifold in an
other way using projective symmetry to give gauge charges for superfields.

\subsection{Implementing global projective symmetry}

\qquad By help of conifold isometries, one may use the gauge transformation
eq(\ref{3}) to go to the matrix coordinate frame where the $2\times 2$
matrix coordinate $Z$ is split into the product of two matrices $Y$ and $X$
as shown below, 
\begin{equation}
Z=YX,
\end{equation}%
with 
\begin{equation}
X=\left( 
\begin{array}{cc}
x & z \\ 
y & w%
\end{array}%
\right) ,\qquad Y=\left( 
\begin{array}{cc}
\sigma & 0 \\ 
0 & \frac{1}{\sigma }%
\end{array}%
\right) ,\qquad \sigma \in \mathbb{C}^{\ast }  \label{6}
\end{equation}%
where, instead of the four complex variables $x_{1},$ $x_{2},$ $y_{1}$ and $%
y_{2}$, we have now five projective complex holomorphic variables $\left(
\sigma ,x,y,z,w\right) $ related to the previous ones like $x=x\left( \sigma
,x_{1}\right) ,$ $y=x\left( \sigma ,y_{1}\right) ,$ $z=z\left( \sigma
,z_{1}\right) ,$ $w=w\left( \sigma ,w_{1}\right) $ as shown below,%
\begin{eqnarray}
x &=&\frac{x_{1}}{\sigma },\qquad z=\frac{x_{2}}{\sigma },  \notag \\
y &=&\sigma y_{1},\qquad x=\sigma y_{2}.  \label{60}
\end{eqnarray}%
In these relations $\sigma \in \mathbb{C}^{\ast }$\ and is a free complex
variable capturing data on the projective abelian sub-symmetry $\mathbb{C}%
^{\ast }$ of the $SL\left( 2,\mathbb{C}\right) $ global isometry of conifold
and where one recognizes the 
\begin{equation}
xy-zw=\mu ,
\end{equation}%
as the defining equation of $T^{\ast }\mathbb{P}^{1}$ geometry embedded in
non compact $W\mathbb{P}^{3}\left( {\small 1,-1,1,-1}\right) $. From the
scaling $\sigma \rightarrow \frac{1}{\lambda }\sigma $, with $\lambda \in 
\mathbb{C}^{\ast }$, we get the projective transformations, 
\begin{equation}
\left( \sigma ,x,y,z,w\right) \rightarrow \left( \sigma ^{\prime },x^{\prime
},y^{\prime },z^{\prime },w^{\prime }\right) =\left( \frac{1}{\lambda }%
\sigma ,\lambda x,\frac{1}{\lambda }y,\lambda z,\frac{1}{\lambda }w\right) ,
\label{9}
\end{equation}%
of the non compact space $\mathbb{WP}^{4}\left( {\small -1,1,-1,1,-1}\right) 
$. Note that old coordinates $x_{1},$ $x_{2},$ $y_{1}$ and $y_{2}$ of $%
\mathbb{C}^{4}$ are recovered from $\sigma ,x,y,z$ and $w$ by fixing
projective symmetry which allows to set\ $\sigma =1$ as in eqs(\ref{60}).
Note also that in a rigorous analysis, the Y matrix should be thought of as 
\begin{equation}
Y=\left( 
\begin{array}{cc}
\sigma _{+} & 0 \\ 
0 & \sigma _{-}%
\end{array}%
\right) ,\qquad \sigma _{-}\sigma _{+}=1,
\end{equation}%
and the non compact $W\mathbb{P}^{4}\left( {\small -1,1,-1,1,-1}\right) $ is
enlarged to $W\mathbb{P}^{5}\left( 1,{\small -1,1,-1,1,-1}\right) $. In the $%
2\times 2$ matrix coordinate frame, conifold equation $\det Z=\mu $ gets
mapped to $\det X=\mu $ where, surprisingly there is no apparent dependence
on the $Y$ matrix variable since $\det Y=1$. Note that like before, the $%
SL\left( 2\right) $ gauge transformations leaving stable conifold eq reads
as 
\begin{equation}
YX\rightarrow Y^{\prime }X^{\prime }=M\left( YX\right) M^{-1},  \label{69}
\end{equation}%
or equivalently as $\left( MYM^{-1}\right) \left( MXM^{-1}\right) $ by
inserting the $2\times 2$ matrix identity $I=M^{-1}M$. Besides the
factorisation $Z=YX$, we should also specify the way to deal\ with\ isometry
eq (\ref{69}). In the standard way, one thinks about the transformed matrix
variables $Y^{\prime }$ and $X^{\prime }$ as given by the change $Y^{\prime
}=\left( MYM^{-1}\right) $ and $X^{\prime }=\left( MXM^{-1}\right) $. The
other possibility we will use below is to think about the transformation $%
Y^{\prime }X^{\prime }=MYXM^{-1}$ as associated with the naive change,%
\begin{equation}
Y^{\prime }=\Lambda Y,\qquad X^{\prime }=X\Lambda ^{-1},  \label{7}
\end{equation}%
with $\Lambda $ an $SL\left( 2,\mathbb{C}\right) $ group matrix. To\ fix the
ideas on the meaning of this transformation, let us consider the global $%
\mathbb{C}^{\ast }$ abelian sub-group of the conifold $SL\left( 2,\mathbb{C}%
\right) $ isometry by making the choice,%
\begin{equation}
\Lambda =\left( 
\begin{array}{cc}
\frac{1}{\lambda } & 0 \\ 
0 & \lambda%
\end{array}%
\right) ,\qquad \lambda \in \mathbb{C}^{\ast },  \label{8}
\end{equation}%
and exhibit the meaning of the transformations (\ref{7}) on the complex
holomorphic coordinates $\left( \sigma ,x,y,z,w\right) $. General
expressions of the transformation $\Lambda $ has to do with conifold
diffeomorphisms; they will be discussed later. Restricting $SL\left( 2,%
\mathbb{C}\right) $ isometry to its abelian global part $\mathbb{C}^{\ast }$%
, then putting eq(\ref{8}) back into eq(\ref{7}), we re-discover eq(\ref{9}%
). Therefore the coordinate mapping $Z=YX$ can be interpreted as moving from
the coordinates $\left( x_{1},x_{2},y_{1},y_{2}\right) $ of $\mathbb{C}^{4}$
to the coordinate frame $\left( \sigma ,x,y,z,w\right) $ parameterizing $W%
\mathbb{P}^{4}\left( {\small -1,1,-1,1,-1}\right) $. In this frame, the
conifold is a described by the invariant projective hypersurface,%
\begin{equation}
F\left( \sigma ,x,y,z,w\right) =xy-zw=\mu ,\qquad \text{and \qquad }\sigma 
\text{ a free }\mathbb{C}^{\ast }\text{ variable,}  \label{10}
\end{equation}%
defining a $T^{\ast }\mathbb{P}^{1}$ fibration over $\mathbb{C}^{\ast }$,
where $T^{\ast }\mathbb{P}^{1}$ is the complex two dimension cotangeant
bundle on complex one dimension projective space $\mathbb{P}^{1}$. In this
fibration, which is also equivalent to eq(\ref{pl}), the holomorphic
variable $\sigma $ parameterizes the base $\mathbb{C}^{\ast }$ and the other
projective coordinates $\left( x,y,z,w\right) $ parameterize the fiber $%
T^{\ast }\mathbb{P}^{1}$. In this view, conifold is given by a projective
hypersurface embedded in $W\mathbb{P}^{4}\left( {\small -1,1,-1,1,-1}\right) 
$.

\qquad The power of this way of doing is that it offers a natural method to
deal with projective functions $\mathcal{G}\left( \sigma ,x,y,z,w\right) $
living on conifold and its sub-manifolds. For projective invariant functions 
$\mathcal{G}\left( \sigma ,x,y,z,w\right) $ on conifold, 
\begin{equation}
\mathcal{G}\left( \frac{1}{\lambda }\sigma ,\lambda x,\frac{1}{\lambda }%
y,\lambda z,\frac{1}{\lambda }w\right) =\mathcal{G}\left( \sigma
,x,y,z,w\right) ,
\end{equation}%
we have two kinds of expansions: (1) a first expansion given by the usual
Laurent development on the base $\mathbb{C}^{\ast }$, 
\begin{equation}
\mathcal{G}\left( \sigma ,x,y,z,w\right) =\sum_{n=-\infty }^{\infty }\sigma
^{n}G_{n}\left( x,y,z,w\right) ,  \label{11}
\end{equation}%
with Laurent modes,%
\begin{equation}
G_{\mp n}\left( x,y,z,w\right) =\frac{1}{2i\pi }\int_{C_{0}}\frac{d\sigma }{%
\sigma ^{\pm n+1}}\mathcal{G}\left( \sigma ,x,y,z,w\right) ,  \label{12}
\end{equation}%
where $C_{0}$ is a contour integral surrounding the pole singularity $\sigma
=0$. (2) Viewed from the fiber sub-manifold, the modes $G_{\mp n}\left(
x,y,z,w\right) $ are projective functions on $T^{\ast }\mathbb{P}^{1}$ with
an integer degree obeying the homogeneity property,%
\begin{equation}
G_{\pm n}\left( \lambda x,\frac{1}{\lambda }y,\lambda z,\frac{1}{\lambda }%
w\right) =\lambda ^{\pm n}G_{\pm n}\left( x,y,z,w\right) .  \label{13}
\end{equation}%
These are just spin $\left( \left\vert n\right\vert +1,0\right) $ and spin $%
\left( 0,\left\vert n\right\vert +1\right) $ representations of the $%
SL\left( 2,C\right) $ isometry group. So they may be expanded in a harmonic
series involving $SL\left( 2\right) $ homogeneous polynomials. For later
use, let us give some details here below; for a complete harmonic space
analysis see $\cite{29}$.

\subsection{Super QFT$_{4}$ realization}

\qquad As we will not have the occasion to discuss in details the super QFT$%
_{4}$ realization of this projective symmetry in forthcoming sections, let
us take this opportunity to fix\ the ideas by giving a superfield
theoretical model realizing this conifold projective geometry. The simplest
model one can imagine is given by a $U_{gauge}\left( 1\right) $
supersymmetric gauge theory with a global $SU\left( 2\right) $ R-symmetry.
The superfield degrees of freedom involved in this gauge theory are reported
on following table. They carry quantum numbers $\left( q,r\right) $
indicating representations.of the $U_{gauge}\left( 1\right) \times
SU_{global}\left( 2\right) $ symmetry.%
\begin{equation}
\begin{tabular}{|l|l|}
\hline
4D $\mathcal{N}=1$ Superfields & $\left( q,r\right) $ Representations \\ 
\hline
$V=-\theta \sigma ^{\mu }\overline{\theta }A_{\mu }-i\overline{\theta }%
^{2}\theta \lambda +i\theta ^{2}\overline{\theta }\overline{\lambda }+\frac{1%
}{2}\theta ^{2}\overline{\theta }^{2}D$ & $\left( 0,1\right) $ \ real gauge
multiplet \\ \hline
$\Phi =\mathrm{\phi }+\theta \mathrm{\psi }+\theta ^{2}\mathrm{F},$ & $%
\left( 0,1\right) $ \ adjoint matter multiplet \\ \hline
$Q_{+\alpha }=q_{\alpha }+\theta \chi _{\alpha }+\theta ^{2}F_{\alpha }$ & $%
\left( 1,2\right) $ fundamental matter \\ \hline
$P_{-\beta }=p_{\beta }+\theta \varphi _{\beta }+\theta ^{2}G_{\beta }$ & $%
\left( -1,\overline{2}\right) $ fundamental matter \\ \hline
$\Sigma _{\pm }=\sigma _{\pm }+\theta \eta _{\pm }+\theta ^{2}L_{\pm }$ & $%
\left( \pm 1,1\right) $ fundamental matter \\ \hline
$\Psi =\gamma _{0}+\theta \tau _{0}+\theta ^{2}G$ & $\left( 0,1\right) $ \
auxiliary superfield \\ \hline
\end{tabular}
\label{tab}
\end{equation}%
where the indices $0,$ $+$ and $-$ carried by $\Sigma _{\pm }$ singlets and
the $SU\left( 2\right) $ superfield doublets $Q_{+\alpha }$ and $P_{-\alpha
} $ refer respectively to the charges of the $U\left( 1\right) $ gauge
group. For convenience, it is interesting to split the superfields $%
Q_{+\alpha }$ and $P_{-\alpha }$ as follows,%
\begin{eqnarray}
\text{{\small hypermultiplet \# 1}} &\text{:}&{\small \qquad }Q_{+\alpha
}=\left( Q_{+,+}\text{ \ },\text{ \ }Q_{+,-}\right) =\left( X_{+}\text{ \ },%
\text{ \ }Z_{+}\right) ,  \notag \\
\text{{\small hypermultiplet \# 2}} &\text{:}&{\small \qquad }P_{-\alpha
}=\left( P_{-,+}\text{ \ },\text{ \ }P_{-,-}\right) =\left( Y_{-}\text{ \ },%
\text{ \ }W_{-}\right) .
\end{eqnarray}%
Then introduce the following neutral chiral superfield doublets to be
considered later,%
\begin{eqnarray}
\text{{\small hypermultiplet \# 3}} &\text{:}&{\small \qquad }H_{0\alpha
}=\left( H_{0,+}\text{ \ },\text{ \ }Q_{0,-}\right) =\left( X_{1}\text{ \ },%
\text{ \ }X_{2}\right)  \notag \\
\text{{\small hypermultiplet \# 4}} &\text{:}&{\small \qquad }K_{0\alpha
}=\left( K_{0,+}\text{ \ },\text{ \ }K_{0,-}\right) =\left( Y_{1}\text{ \ },%
\text{ \ }Y_{2}\right) .
\end{eqnarray}%
The superspace lagrangian density $\mathcal{L}=\mathcal{L}\left( T^{\ast }%
\mathbb{S}^{3}\right) $ describing the dynamics of these superfields and
preserving $U_{gauge}\left( 1\right) \times SU_{global}\left( 2\right) $
symmetry may be split as 
\begin{equation}
\mathcal{L}=\mathcal{L}_{1}+\mathcal{L}_{2}+\left( \mathcal{L}_{3}+hc\right)
.
\end{equation}%
The first term reads as,%
\begin{equation}
\mathcal{L}_{1}=\mathcal{L}_{g}\left( V\right) +\mathcal{L}_{ad}\left( \Phi
\right) -2\zeta \int d^{4}\theta V-\left( \mu \int d^{2}\theta \Phi
+hc\right) ,
\end{equation}%
where $\mathcal{L}_{g}\left( V\right) $ and $\mathcal{L}_{ad}\left( \Phi
\right) $ stand respectively for the usual gauge covariant lagrangian
densities of the $U\left( 1\right) $ vector multiplet and adjoint matter
superfield and where the parameters $\zeta $ and $\mu $ are the usual Fayet
Iliopoulos coupling constants. The second term is given by the usual gauge
covariant kinetic terms, 
\begin{equation}
\mathcal{L}_{2}=\int d^{4}\theta \sum_{\alpha =1}^{2}\left( \overline{\left(
Q_{+\alpha }\right) }e^{2V}Q_{+\alpha }+\overline{\left( P_{-\alpha }\right) 
}e^{-2V}P_{-\alpha }\right) +\int d^{4}\theta \overline{\left( \Sigma _{\pm
}\right) }e^{-2V}\Sigma _{\pm }.
\end{equation}%
The third term $\mathcal{L}_{3}=\mathcal{L}_{3}\left( \Psi ,\Phi ,Q_{+\alpha
},P_{-\beta },\Sigma _{\pm }\right) $ deals with the chiral and antichiral
superpotential. The chiral sector factor reads as follows,%
\begin{eqnarray}
\mathcal{L}_{3} &=&\int d^{2}\theta \left( g_{0}\Psi \left( \Sigma
_{+}\Sigma _{-}-1\right) \right)  \notag \\
&&+\int d^{2}\theta \left( g_{1}\Phi Q_{+\alpha }P_{-\beta }\varepsilon
^{\alpha \beta }-g_{2}\Phi \Sigma _{+}\Sigma \right) \\
&&+\int d^{2}\theta \mathrm{W}\left( Q,P,\Sigma \right) ,  \notag
\end{eqnarray}%
where the g$_{i}$'s are coupling constants. Note that eliminating \ the
auxiliary superfields $\Psi $ and $\Phi $ through their holomorphic eqs of
motion, one gets, 
\begin{eqnarray}
\Sigma _{+}\Sigma _{-} &=&1,  \notag \\
g_{1}Q_{+\alpha }P_{-\beta }\varepsilon ^{\alpha \beta } &=&\mu +g_{2}\Sigma
_{+}\Sigma .
\end{eqnarray}%
We also recall the following useful relations,%
\begin{eqnarray}
Q_{+\alpha }P_{-\beta }\varepsilon ^{\alpha \beta } &=&X_{+}Y_{-}-Z_{+}W_{-},
\notag \\
H_{0\alpha }K_{0\beta }\varepsilon ^{\alpha \beta } &=&X_{1}Y_{2}-X_{2}Y_{1},
\label{id} \\
Q_{+\alpha }Q_{+\beta }\varepsilon ^{\alpha \beta } &=&P_{-\alpha }P_{-\beta
}\varepsilon ^{\alpha \beta }=H_{0\alpha }H_{0\beta }\varepsilon ^{\alpha
\beta }=0.  \notag
\end{eqnarray}%
Substituting $\Sigma _{+}\Sigma _{-}=1$ in the first relation and shifting $%
\mu \rightarrow g_{1}\left( \mu -g_{2}\right) $, we discover $Q_{+\alpha
}P_{-\beta }\varepsilon ^{\alpha \beta }=\mu $. Setting $P_{-\alpha
}=K_{0\alpha }\Sigma _{-}$ and $H_{0\alpha }=\Sigma _{-}Q_{+\alpha }$, we
end with,%
\begin{equation}
H_{0\alpha }K_{0\beta }\varepsilon ^{\alpha \beta }=\mu ,
\end{equation}%
which, by help of the identity (\ref{id})\ is nothing but the conifold eq $%
X_{1}Y_{2}-X_{2}Y_{1}=\mu $ in superfield language.

\section{Diffeomorphisms}

\qquad\ In the old coordinate system $\left\{ x_{i},y_{j}\right\} $\ where
conifold is seen as hypersurface $\varepsilon ^{ij}x_{i}y_{j}=\mu $ embedded
in $\mathbb{C}^{4}$ eqs(\ref{4}-\ref{5}), conifold isometries are given by
the general coordinate transformations 
\begin{eqnarray}
x_{i}^{\prime } &=&x_{i}^{\prime }\left( x,y\right) ,  \notag \\
y_{i}^{\prime } &=&y_{i}^{\prime }\left( x,y\right) ,  \label{51}
\end{eqnarray}%
leaving $\varepsilon ^{ij}x_{i}y_{j}$ invariant. Since $x_{i}$\ and $y_{j}$\
are rotated under $SL\left( 2\right) $, it is not difficult to see that
these general coordinate transformations should be as,%
\begin{eqnarray}
x_{i}^{\prime } &=&y_{k}\Upsilon _{i}^{k},\qquad \Upsilon _{i}^{k}=\Upsilon
_{i}^{k}\left( x,y\right) ,  \notag \\
y_{j} &=&\Gamma _{j}^{l}x_{l}\qquad \Gamma _{j}^{l}=\Gamma _{j}^{l}\left(
x,y\right) ,
\end{eqnarray}%
where the local matrices $\Upsilon _{i}^{k}$\ and $\Gamma _{j}^{l}$\ are
constrained as 
\begin{equation}
\varepsilon ^{ij}\Upsilon _{i}^{k}\Gamma _{j}^{l}=1,
\end{equation}%
showing $\Gamma _{j}^{l}$\ is the inverse of $\Upsilon _{i}^{k}$\ and $\det
\Upsilon =1$. Global matrices $\Upsilon $ generate then the global $SL\left(
2\right) $ sub-symmetry of $diff\left( T^{\ast }S^{3}\right) $.

\qquad In the projective coordinate frame $\left( \sigma ,x,y,z,w\right) $
of $W\mathbb{P}^{4}\left( {\small -1,1,-1,1,-1}\right) $ eqs(\ref{60}-\ref{9}%
), we have a quite similar description, except that now we have more
explicited transformation from which we can also read the change concerning
the base sub-manifold and fiber. In this frame, conifold diffeomorphisms are
given by general coordinate transformations,%
\begin{eqnarray}
\sigma ^{\prime } &=&\sigma ^{\prime }\left( \sigma ,x,y,z,w\right) ,  \notag
\\
x^{\prime } &=&x^{\prime }\left( \sigma ,x,y,z,w\right) ,  \notag \\
y^{\prime } &=&y^{\prime }\left( \sigma ,x,y,z,w\right) ,  \label{gen} \\
z^{\prime } &=&z^{\prime }\left( \sigma ,x,y,z,w\right) ,  \notag \\
w^{\prime } &=&w^{\prime }\left( \sigma ,x,y,z,w\right) ,  \notag
\end{eqnarray}%
preserving projective symmetry; i.e,%
\begin{eqnarray}
\sigma ^{\prime }\left( \frac{1}{\lambda }\sigma ,\lambda x,\frac{1}{\lambda 
}y,\lambda z,\frac{1}{\lambda }w\right) &=&\frac{1}{\lambda }\sigma ^{\prime
}\left( \sigma ,x,y,z,w\right)  \notag \\
x^{\prime }\left( \frac{1}{\lambda }\sigma ,\lambda x,\frac{1}{\lambda }%
y,\lambda z,\frac{1}{\lambda }w\right) &=&\lambda x^{\prime }\left( \sigma
,x,y,z,w\right)  \notag \\
y^{\prime }\left( \frac{1}{\lambda }\sigma ,\lambda x,\frac{1}{\lambda }%
y,\lambda z,\frac{1}{\lambda }w\right) &=&\frac{1}{\lambda }y^{\prime
}\left( \sigma ,x,y,z,w\right)  \label{gem} \\
z^{\prime }\left( \frac{1}{\lambda }\sigma ,\lambda x,\frac{1}{\lambda }%
y,\lambda z,\frac{1}{\lambda }w\right) &=&\lambda z^{\prime }\left( \sigma
,x,y,z,w\right)  \notag \\
w^{\prime }\left( \frac{1}{\lambda }\sigma ,\lambda x,\frac{1}{\lambda }%
y,\lambda z,\frac{1}{\lambda }w\right) &=&\frac{1}{\lambda }w^{\prime
}\left( \sigma ,x,y,z,w\right) .  \notag
\end{eqnarray}%
Restricting the variable change $\left( x^{\prime },y^{\prime },z^{\prime
},w^{\prime }\right) $ to the special case $x^{\prime }=x^{\prime }\left(
x,y,z,w\right) $ and so on, gives diffeomorphisms of $T^{\ast }\mathbb{P}%
^{1} $. Restriction to the base $\sigma ^{\prime }=\sigma ^{\prime }\left(
\sigma \right) $ gives $\mathbb{C}^{\ast }$ diffeomorphisms. Note that
fixing $\sigma ^{\prime }=\sigma $, the above changes reduce to general
coordinate transformations on $T^{\ast }\mathbb{P}^{1}$ fiber. The same
observations are valid for the $\mathbb{S}^{3}$, $\mathbb{S}^{2}$ and $%
\mathbb{S}^{1}$ real slices. The general coordinates transformations (\ref%
{gen}-\ref{gem}) translate into the corresponding two component projective
isospinor formalism as follows,%
\begin{eqnarray}
u_{i}^{\prime } &=&u_{i}^{\prime }\left( \sigma ,u_{j},v_{j}\right) ,  \notag
\\
v_{i}^{\prime } &=&v_{i}^{\prime }\left( \sigma ,u_{j},v_{j}\right) ,
\end{eqnarray}%
with%
\begin{eqnarray}
u_{i}^{\prime }\left( \frac{1}{\lambda }\sigma ,\lambda u_{j},\frac{1}{%
\lambda }v_{j}\right) &=&\lambda u_{i}^{\prime }\left( \sigma
,u_{j},v_{j}\right) ,  \notag \\
v_{i}^{\prime }\left( \frac{1}{\lambda }\sigma ,\lambda u_{j},\frac{1}{%
\lambda }v_{j}\right) &=&\frac{1}{\lambda }v_{i}^{\prime }\left( \sigma
,u_{j},v_{j}\right) .
\end{eqnarray}%
Notice that here $u_{j}=\left( x,z\right) $ and $v_{j}=\left( y,w\right) $
are projective coordinates; they should not be confused with those given by
eqs(\ref{51}).

The charge operator of the projective symmetry on $\mathbb{WP}^{4}\left( 
{\small -1,1,-1,1,-1}\right) $ namely, 
\begin{equation}
\nabla _{0}=\left( x\frac{\partial }{\partial x}+z\frac{\partial }{\partial z%
}\right) -\left( y\frac{\partial }{\partial y}+w\frac{\partial }{\partial w}%
\right) -\sigma \frac{\partial }{\partial \sigma },
\end{equation}%
splits into two commuting parts as 
\begin{equation}
\nabla _{0}=2D_{0}-T_{0}
\end{equation}%
with contribution $T_{0}$ coming from the base, 
\begin{equation}
T_{0}=\nabla _{0}|_{\mathbb{C}^{\ast }},
\end{equation}%
and a second one $D_{0}$ coming from the fiber, 
\begin{equation}
2D_{0}=\nabla _{0}|_{T^{\ast }\mathbb{P}^{1}}.
\end{equation}%
The charge operator $T_{0}$, which counts the projective charges of sections
along the $\mathbb{C}^{\ast }$ base, reads as, 
\begin{equation}
T_{0}=\sigma \frac{\partial }{\partial \sigma },  \label{14}
\end{equation}%
and has integer eigenvalues $n$ and eigen-functions given by the usual
Laurent monomials $f_{n}\sim \sigma ^{n}$. The $D_{0}$ charge operator deals
with the counting of the Cartan Weyl charges on the $T^{\ast }\mathbb{P}^{1}$
base; it reads then as,%
\begin{equation}
2D_{0}=\left( x\frac{\partial }{\partial x}+z\frac{\partial }{\partial z}%
\right) -\left( y\frac{\partial }{\partial y}+w\frac{\partial }{\partial w}%
\right) .  \label{15}
\end{equation}%
It has integer eigenvalues and eigen-functions given by harmonic functions
on $T^{\ast }\mathbb{P}^{1}$. In this setting, projective invariance of
functions $G=G\left( T^{\ast }S^{3}\right) $ living on $T^{\ast }S^{3}$, is
solved as,%
\begin{equation}
\nabla _{0}G\left( \sigma ,x,y,z,w\right) =0,
\end{equation}%
implying in turns,%
\begin{equation}
2D_{0}G=T_{0}G.  \label{16}
\end{equation}%
This identity means that projective charges on fiber and base sub-manifolds
cancel themselves exactly. Notice that in general, projective covariance on
conifold requires,%
\begin{equation}
eignvalue\left( 2D_{0}\right) -eignvalue\left( T_{0}\right) \text{ \ }\in 
\text{ \ }\mathbb{Z}.
\end{equation}%
On the cotangeant bundle $T^{\ast }\mathbb{P}^{1}$ described by the
projective invariant eq $xy-zw=\mu $ with $x\equiv \lambda x$, $y\equiv 
\frac{1}{\lambda }y$, $z\equiv \lambda z$ and $w\equiv \frac{1}{\lambda }w$
but no $\sigma $ dependence, it happens that $D_{0}$ is in fact one of a set
of three operators namely $D_{0}$ and $D_{\pm }$. These operators generate
the $SL\left( 2,\mathbb{C}\right) $ global sub-group of conifold
diffeomorphism isometries. The commutation relations of $D_{0}$ and $D_{\pm
} $ are given by similar relations to those of $SU\left( 2,\mathbb{C}\right) 
$ except here we have no hermiticity conditions, 
\begin{eqnarray}
\left[ D_{+},D_{-}\right] &=&2D_{0},  \notag \\
\left[ 2D_{0},D_{\pm }\right] &=&\pm 2D_{\pm }.  \label{17}
\end{eqnarray}%
Together with the expression of $2D_{0}$\ given above eq(\ref{15}), the $%
D_{\pm }$ operators realizing the above $SL\left( 2,\mathbb{C}\right) $
brackets read as,%
\begin{eqnarray}
D_{+} &=&x\frac{\partial }{\partial w}-z\frac{\partial }{\partial y},  \notag
\\
D_{-} &=&w\frac{\partial }{\partial x}-y\frac{\partial }{\partial z}.
\label{18}
\end{eqnarray}%
Under this realization of $SL\left( 2,\mathbb{C}\right) $ global isometry,
one recovers the results described above. In particular we recover that the
projective variables $u=\left( x,z\right) $ and $v=\left( y,w\right) $ carry
respectively the projective charges $\left( +,+\right) $ and $\left(
-,-\right) $ and moreover transform as $SL\left( 2,\mathbb{C}\right) $
isodoublets as shown below,%
\begin{equation}
\left[ \nabla _{0},\left( x,z\right) \right] =\left( x,z\right) ,\qquad %
\left[ \nabla _{0},\left( y,w\right) \right] =\left( -y,-z\right) ,
\label{19}
\end{equation}%
and 
\begin{eqnarray}
\left[ \nabla _{+},\left( x,z\right) \right] &=&\left( 0,0\right)  \notag \\
\left[ \nabla _{-},\left( x,z\right) \right] &=&\left( w,-y\right) ,
\label{20} \\
\left[ \nabla _{-}^{2},\left( x,z\right) \right] &=&\left( 0,0\right) . 
\notag
\end{eqnarray}%
Similar relations for $y$ and $w$ may be also written down; in particular we
have $\left[ \nabla _{-},\left( y,w\right) \right] =\left( 0,0\right) $. To
complete the picture let us make three comments concerning diffeomorphism
isometries of the conifold.

First note that in the projective frame, there is a one to one
correspondence between base and fiber objects,%
\begin{equation}
\begin{tabular}{|l|l|l|l|l|}
\hline
$T^{\ast }\mathbb{P}^{1}$ fiber & $\leftrightarrow $ & $\mathbb{C}^{\ast }$
base & $\leftrightarrow $ & Conifold \\ \hline
$G_{n}=G_{n}\left( x,y,z,w\right) $ &  & $f_{n}=\sigma ^{n}$ &  & $G\left(
\sigma ,x,y,z,w\right) =\sum \sigma ^{n}G_{n}$ \\ \hline
$D_{+}=x\frac{\partial }{\partial w}-z\frac{\partial }{\partial y}$ &  & $%
T_{+}=\frac{1}{\sqrt{2}}\frac{\partial }{\sigma \partial \sigma }$ &  & $%
\nabla _{+}=D_{+}-T_{+}$ \\ \hline
$2D_{0}=\frac{x\partial }{\partial x}+\frac{z\partial }{\partial z}-\frac{%
y\partial }{\partial y}-\frac{w\partial }{\partial w}$ &  & $T_{0}=\sigma 
\frac{\partial }{\partial \sigma }$ &  & $\nabla _{0}=2D_{0}-T_{0}$ \\ \hline
$D_{-}=w\frac{\partial }{\partial x}-y\frac{\partial }{\partial z}$ &  & $%
T_{-}=-\frac{\sigma ^{3}\partial }{\sqrt{2}\partial \sigma }$ &  & $\nabla
_{-}=D_{-}-T_{-}$ \\ \hline
$2D_{0}G_{n}=nG_{n}$ &  & $T_{0}f_{n}=nf_{n}$ &  & $\nabla _{0}G=0$ \\ \hline
\end{tabular}%
,
\end{equation}%
where $n\in Z$. Second, like $D_{0}$ and $D_{\pm }$, the generators $\left(
-T_{0}\right) $ and $\left( -T_{\pm }\right) $ obey an $SL\left( 2,C\right) $
algebra. In addition to eq(\ref{14}), the realization of the generators $%
T_{\pm }$ is given by, 
\begin{equation}
T_{+}=\frac{1}{\sqrt{2}}\frac{\partial }{\sigma \partial \sigma },\qquad
T_{-}=-\frac{1}{\sqrt{2}}\frac{\sigma ^{3}\partial }{\partial \sigma }.
\end{equation}%
It is dictated by the projective transformation $\sigma \rightarrow \frac{1}{%
\lambda }\sigma $ acting on $T_{\pm }$ as $\lambda ^{\pm 2}T_{\pm }$ that is
in same manner as does the change $\left( x,y,z,w\right) \rightarrow \left(
\lambda x,\frac{1}{\lambda }y,\lambda z,\frac{1}{\lambda }w\right) $ on the $%
D_{\pm }$ generators of the $T^{\ast }\mathbb{P}^{1}$ base. It follows also
from the representation%
\begin{equation}
T_{+}=\frac{1}{\sqrt{2}}\sigma _{+}\frac{\partial }{\partial \sigma _{-}}%
,\qquad T_{-}=\frac{1}{\sqrt{2}}\sigma _{-}\frac{\partial }{\partial \sigma
_{+}},\qquad T_{0}=\frac{1}{2}\left( \sigma _{+}\frac{\partial }{\partial
\sigma _{+}}-\sigma _{-}\frac{\partial }{\partial \sigma _{-}}\right) ,
\end{equation}%
and substituting $\sigma _{+}\sigma _{-}=1$. Finally under global projective
symmetry, the $SL\left( 2\right) $ generators of conifold scale in same
manner as for $D_{q}$\ and $T_{q}$\ parts namely,%
\begin{equation}
\nabla _{q}\rightarrow \lambda ^{2q}\nabla _{q},  \label{hh}
\end{equation}%
with $q=0,\pm 1.$

\subsection{Conifold isometries}

\qquad First note that, along with the particular global projective symmetry
described above (\ref{9}), conifold has an infinite set of diffeomorphism
symmetries sitting in the group $Diff\left( T^{\ast }\mathbb{S}^{3}\right) $%
. Roughly speaking, and as far as the $T^{\ast }\mathbb{P}^{1}$ fiber is
concerned, there are three main subsets which read in the $T^{\ast }\mathbb{P%
}^{1}$ projective isospinor formalism as,%
\begin{equation}
\left( 
\begin{array}{c}
u \\ 
v%
\end{array}%
\right) \rightarrow M\left( 
\begin{array}{c}
u \\ 
v%
\end{array}%
\right) ,
\end{equation}%
with matrix $M$ as follows,%
\begin{equation}
M_{0}=\left( 
\begin{array}{cc}
\Lambda & 0 \\ 
0 & \Lambda ^{-1}%
\end{array}%
\right) ,\qquad M_{1}=\left( 
\begin{array}{cc}
1 & \Upsilon \\ 
0 & 1%
\end{array}%
\right) ,\qquad M_{2}=\left( 
\begin{array}{cc}
1 & 0 \\ 
\Gamma & 1%
\end{array}%
\right) .
\end{equation}%
where $\Lambda \neq 0$, $\Upsilon $ and $\Gamma $ are arbitrary functions on
conifold. To complete the picture, one should also add the general
coordinate transformation of the base $\sigma ^{\prime }=\sigma ^{\prime
}\left( \sigma ,u,v\right) $; it will be implemented later. A particular
subset of $u$ and $v$ general coordinate transformations is that given by
the holomorphic change $M_{1}$ mapping half of the $T^{\ast }\mathbb{P}^{1}$
projective coordinate, say $x$ and $z$, as 
\begin{eqnarray}
x &\rightarrow &x^{\prime }\left( \sigma ,x,y,z,w\right) =x+\varepsilon
\left( \sigma ,x,y,z,w\right) ,  \notag \\
z &\rightarrow &z^{\prime }\left( \sigma ,x,y,z,w\right) =z+\epsilon \left(
\sigma ,x,y,z,w\right) ,  \label{210}
\end{eqnarray}%
and fixing the other half since leaving the variables $y$ and $w$ unchanged,%
\begin{equation}
y\rightarrow y^{\prime }=y,\qquad \qquad w\rightarrow w^{\prime }=w.
\end{equation}%
In the projective coordinate frame we are working with here, explicit
general coordinates transformations that leave invariant conifold
hypersurface $xy-zw=\mu $ read as,%
\begin{eqnarray}
x &\rightarrow &x^{\prime }=x+\Upsilon w,  \notag \\
z &\rightarrow &z^{\prime }=z+\Upsilon y,  \label{21} \\
y &\rightarrow &y^{\prime }=y,\qquad w\rightarrow w^{\prime }=w.  \notag
\end{eqnarray}%
If focusing on the $T^{\ast }\mathbb{P}^{1}$ coordinates where projective
symmetry should be imposed, the general coordinates parameter $\Upsilon
=\Upsilon \left( \sigma ,x,y,z,w\right) $ is an arbitrary degree two
homogeneous function on conifold,%
\begin{equation}
\Upsilon \left( \frac{1}{\lambda }\sigma ,\lambda x,\frac{1}{\lambda }%
y,\lambda z,\frac{1}{\lambda }w\right) =\lambda ^{2}\Upsilon \left( \sigma
,x,y,z,w\right) .  \label{22}
\end{equation}%
This is because of the opposite projective charges of $\left( x,z\right) $
and $\left( y,w\right) $. According to the analysis of previous section,
this function expands in a Laurent series as follows,%
\begin{equation}
\Upsilon \left( \sigma ,x,y,z,w\right) =\sum_{n=-\infty }^{\infty }\sigma
^{n}\Upsilon _{n+2}\left( x,y,z,w\right) ,
\end{equation}%
with 
\begin{equation}
\Upsilon _{n+2}=\doint \frac{d\sigma }{2i\pi \sigma ^{n+1}}\Upsilon ,
\end{equation}%
being the Laurent modes and at same time are functions living on $T^{\ast }%
\mathbb{P}^{1}$. Recall that $x$ and $z$ have projective degree one and $%
\sigma ,$ $y$ and $w$ have a degree minus one. The constraint eq(\ref{22}),
which reads also as 
\begin{equation}
\left[ 2\nabla _{0},\Upsilon \right] =2\Upsilon ,
\end{equation}%
or equivalently by using the splitting $2\nabla _{0}=2D_{0}-T_{0}$, 
\begin{eqnarray}
\left[ 2D_{0},\sigma ^{n}\Upsilon _{n+2}\right] &=&\left( n+2\right)
\Upsilon _{n+2},  \notag \\
\left[ T_{0},\sigma ^{n}\Upsilon _{n+2}\right] &=&n\Upsilon _{n+2},
\end{eqnarray}%
has infinitely many solutions for $\Upsilon _{n+2}$ classified by $SL\left(
2,\mathbb{C}\right) $ spins $\left( s_{1},s_{2}\right) $. The two simplest
examples read respectively as 
\begin{equation}
\Upsilon =\frac{1}{\sigma ^{2}},\qquad D_{0,\pm }\left( \frac{1}{\sigma ^{2}}%
\right) =0,
\end{equation}%
living on base; i.e no dependence on fiber coordinate variables, and 
\begin{equation}
\Upsilon =\frac{ax+bz}{\sigma },\qquad 2D_{0}\left( \frac{ax+bz}{\sigma }%
\right) =-T_{0}\left( \frac{ax+bz}{\sigma }\right) =\frac{ax+bz}{\sigma },
\end{equation}%
with a foot in the base and the other in the fiber. The next example coming
after is given by the following isotriplet representation $\left( 1,0\right) 
$ living in the base,%
\begin{equation}
\Upsilon \equiv \Upsilon _{2}=\left( ax^{2}+bxz+cz^{2}\right) ,\qquad
T_{0,\pm }\left( \Upsilon _{2}\right) =0,  \label{221}
\end{equation}%
with $a$, $b$ and $c$ are complex parameters. Form eq(\ref{15}), one can
easily check that the property $\left[ 2D_{0}-T_{0},\Upsilon \right]
=2\Upsilon $ is fulfilled; and by using eq(\ref{18}), one finds that it
satisfies moreover $\left[ D_{+},\Upsilon _{2}\right] =0$ showing that $%
\Upsilon _{2}$ is indeed a highest weight state of spin $\left( 1,0\right) $.

\qquad Together with the general coordinate transformations (\ref{210}), we
have also a mirror set of diffeomorphisms fixing the isodoublet $\left(
x,z\right) $, that is $x\rightarrow x^{\prime }=x,$ $z\rightarrow z^{\prime
}=z$, but changing the second isodoublet $\left( y,w\right) $ as follows,%
\begin{eqnarray}
y &\rightarrow &y^{\prime }=y+\Gamma z,  \notag \\
w &\rightarrow &w^{\prime }=w+\Gamma x.  \label{23}
\end{eqnarray}%
Here also, the diffeomorphism group parameter $\Gamma =\Gamma \left( \sigma
,x,y,z,w\right) $ is given by homogeneous function living on conifold and
has a degree $\left( -2\right) $, 
\begin{equation}
\Gamma \left( \frac{1}{\lambda }\sigma ,\lambda x,\frac{1}{\lambda }%
y,\lambda z,\frac{1}{\lambda }w\right) =\lambda ^{-2}\Gamma \left( \sigma
,x,y,z,w\right) .  \label{24}
\end{equation}%
Like before, this holomorphic function expands in a Laurent series as $%
\Gamma \left( \sigma ,x,y,z,w\right) =\sum_{n=-\infty }^{\infty }\sigma
^{n}\Gamma _{n-2}\left( x,y,z,w\right) $. In the $SL\left( 2,\mathbb{C}%
\right) $ differential operator language, the condition (\ref{24}) maps to,%
\begin{equation}
\left[ 2D_{0},\Gamma _{n-2}\right] =\left( n-2\right) \Gamma _{n-2},
\label{241}
\end{equation}%
and has infinitely many solutions. The simplest three solutions read
respectively as $\Gamma =\sigma ^{2}$ living on base, $\Gamma =\sigma \left(
ay+bw\right) $ with a foot in fiber and the other in base and the third one
is given by the $\left( 0,1\right) $ isotriplet representation,%
\begin{equation}
\Gamma =\left( ay^{2}+byw+cz^{2}\right) ,  \label{25}
\end{equation}%
with $a$, $b$ and $c$ some arbitrary group parameters. It lives in the $%
T^{\ast }\mathbb{P}^{1}$ fiber sub-manifold.

\subsection{Local $\mathbb{C}^{\ast }$ symmetry}

\qquad Here we complete the previous analysis by considering the study of
local projective symmetry (\ref{9}). This concerns the abelian gauge
sub-symmetry obtained by setting $\Upsilon =0$, $\Gamma =0$ and $\digamma =0$
in the following typical general coordinate transformations. Recall that the
general coordinates transformations in u-sector reads as,%
\begin{eqnarray}
x &\rightarrow &x^{\prime }=\Lambda x+\Lambda \Upsilon w,  \notag \\
z &\rightarrow &z^{\prime }=\Lambda z+\Lambda \Upsilon y,  \notag \\
y &\rightarrow &y^{\prime }=\frac{1}{\Lambda }y,\qquad w\rightarrow
w^{\prime }=\frac{1}{\Lambda }w, \\
\sigma &\rightarrow &\sigma ^{\prime }=\frac{1}{\Lambda }\left( \sigma
+\digamma \right) ,  \notag
\end{eqnarray}%
with gauge parameters $\Lambda =\Lambda \left( \sigma ,x,y,z,w\right) $, $%
\Upsilon =\Upsilon \left( \sigma ,x,y,z,w\right) $ and $\digamma =\digamma
\left( \sigma ,x,y,z,w\right) $ while their analog for $v$-sector have the
form, 
\begin{eqnarray}
x &\rightarrow &x^{\prime }=\Lambda x,\qquad z\rightarrow z^{\prime
}=\Lambda z,  \notag \\
\sigma &\rightarrow &\sigma ^{\prime }=\frac{1}{\Lambda }\left( \sigma
+\digamma \right) ,  \notag \\
y &\rightarrow &y^{\prime }=\frac{1}{\Lambda }y+\frac{\Gamma }{\Lambda }z, \\
w &\rightarrow &w^{\prime }=\frac{1}{\Lambda }w+\frac{\Gamma }{\Lambda }x, 
\notag
\end{eqnarray}
As noted previously, since the defining equation of the conifold $\det
\left( YX\right) =\mu $ involves no coordinate derivatives, the projective
change 
\begin{equation}
Y^{\prime }=\mathbf{\Lambda }Y,\qquad X^{\prime }=X\mathbf{\Lambda }^{-1},
\end{equation}%
with $\mathbf{\Lambda }$\textrm{\ }as in eq(\ref{8}), is also valid for a
local group parameter 
\begin{equation}
\Lambda =\Lambda \left( \sigma ,x,y,z,w\right) ,
\end{equation}%
living on conifold. To fix the ideas, $\Lambda $ may be thought of as given
by, 
\begin{equation}
\Lambda \left( \sigma ,x,y,z,w\right) =\lambda \exp \left[ \eta \left(
\sigma ,x,y,z,w\right) \right] ,  \label{26}
\end{equation}%
where the non zero complex constant $\lambda $ is as before and where $\eta
=\eta \left( \sigma ,x,y,z,w\right) $ is an arbitrary local projective
function. Like before, a class of these functions $\Lambda $ is given by
homogeneous a function of degree zero; i.e,%
\begin{equation}
\Lambda \left( \frac{1}{\lambda }\sigma ,\lambda x,\frac{1}{\lambda }%
y,\lambda z,\frac{1}{\lambda }w\right) =\Lambda \left( \sigma
,x,y,z,w\right) ,
\end{equation}%
Conservation of the global projective charge shows that this kind of
function $\Lambda $ may be expanded in a Laurent series as follows,%
\begin{equation}
\Lambda \left( \sigma ,x,y,z,w\right) =\sum_{n=-\infty }^{\infty }\sigma
^{n}\Lambda _{n},
\end{equation}%
where $\Lambda _{n}=\Lambda _{n}\left( x,y,z,w\right) $ are projective
functions of order $n$ living on $T^{\ast }\mathbb{P}^{1}$ and satisfying,%
\begin{equation}
\left[ 2D_{0},\Lambda _{n}\right] =n\Lambda _{n}
\end{equation}%
Like before, there are infinitely many solutions classified by $SL\left( 2,%
\mathbb{C}\right) $ representations. The simplest one is naturally the\
global constant $\Lambda _{0}=\lambda $ of spin $\left( s_{1},s_{1}\right)
=\left( 0,0\right) $ while the two next ones read as 
\begin{equation}
\Lambda =\sigma \left( ax+bz\right) ,
\end{equation}%
and 
\begin{equation}
\Lambda =\frac{ay+bw}{\sigma },
\end{equation}%
and are respectively associated with $\left( \frac{1}{2},0\right) $ and $%
\left( 0,\frac{1}{2}\right) $\ representations of $SL\left( 2\right) $
global isometry group. The next solution, which is given by the spin $\left( 
\frac{1}{2},\frac{1}{2}\right) $, reads as 
\begin{equation}
\Lambda =axw+bzy+c\left( xy+zw\right) ,  \label{27}
\end{equation}%
with $a$, $b$ and $c$ are arbitrary complex parameters. As such the global
projective eqs(\ref{8}) extends locally as 
\begin{eqnarray}
x^{\prime } &=&\Lambda x,\qquad z^{\prime }=\Lambda z,  \notag \\
y^{\prime } &=&\frac{1}{\Lambda }y,\qquad w^{\prime }=\frac{1}{\Lambda }w,
\label{28}
\end{eqnarray}%
and is interpreted as a $\mathbb{C}^{\ast }$ gauge symmetry acting on
projective functions living on conifold. In next section, we fix our
attention on this abelian local symmetry and look for the corresponding
underlying gauge theory.

\section{More on gauging $\mathbb{C}^{\ast }$ isometry}

\qquad To start recall that in the projective transformation (\ref{9}), the
parameter $\lambda $ is a non zero global $SL\left( 2,\mathbb{C}\right) $
scalar; that is a non zero complex constant satisfying 
\begin{equation}
\left[ D_{0},\lambda \right] =\left[ D_{\pm },\lambda \right] =0,\qquad %
\left[ T_{0},\lambda \right] =\left[ T_{\pm },\lambda \right] =0,
\end{equation}%
that is%
\begin{equation}
\left[ \nabla _{0},\lambda \right] =\left[ \nabla _{\pm },\lambda \right] =0.
\end{equation}%
Under this global $\mathbb{C}^{\ast }$ symmetry, generic field sections $%
G_{n}=G_{n}\left( x,y,z,w\right) $ with $n$ charges and their derivatives $%
D_{0,\pm }G_{n}$ transform covariantly as in eq(\ref{13}) namely 
\begin{equation}
G_{n}\rightarrow \lambda ^{n}G_{n},\qquad \left( D_{0,\pm }G_{n}\right)
\rightarrow \lambda ^{n}\left( D_{0,\pm }G_{n}\right) .
\end{equation}%
The same transformations are valid for the global $SL\left( 2,\mathbb{C}%
\right) $ generators $D_{0,\pm }$ which transform then as,%
\begin{equation}
D_{\pm }\rightarrow D_{\pm }^{\prime }=\lambda ^{\pm 2}D_{\pm },\qquad
2D_{0}\rightarrow 2D_{0}^{\prime }=2D_{0}.
\end{equation}%
Similar relations may be written down for the analogous base quantities, for
instance $\sigma ^{n}\rightarrow \lambda ^{-n}\sigma ^{n}$ and $%
T_{0}\rightarrow T_{0}^{\prime }=T_{0}$, $T_{\pm }\rightarrow T_{\pm
}^{\prime }=\lambda ^{\pm 2}T_{\pm }$. The same is valid for $\nabla _{0,\pm
}$, eq(\ref{hh}). Extending the global projective symmetry of parameter $%
\lambda $ to a local one with arbitrary gauge parameter $\Lambda $ living on
conifold, 
\begin{equation}
\left[ \nabla _{0,\pm },\Lambda \right] \neq 0,
\end{equation}%
the conifold hypersurface $\left\{ \left( \sigma ,x,y,z,w\right) \text{ }|%
\text{ }xy-zw=\mu ;\text{ \ }\sigma \in \mathbb{C}^{\ast }\right\} $ remains
invariant. But what about field on conifold and their derivatives. To that
purpose let us first focus on the field sections $G_{n}\left( x,y,z,w\right) 
$ on $T^{\ast }\mathbb{P}^{1}$. These fields transform covariantly as $%
G_{n}\rightarrow \Lambda ^{n}G_{n}$; however the derivatives $D_{\pm }G_{n}$
and $D_{0}G_{n}$ are no longer covariant since they undergo like,%
\begin{eqnarray}
D_{\pm }G_{n} &=&\Lambda ^{n\pm 2}\left[ D_{\pm }+nD_{\pm }\left( \ln
\Lambda \right) \right] G_{n},  \notag \\
D_{0}G_{n} &=&\Lambda ^{n}\left[ D_{0}+nD_{0}\left( \ln \Lambda \right) %
\right] G_{n}.
\end{eqnarray}%
These are typical transformations in gauge theories which rests on the fact
that the derivatives $D_{0,\pm }$ are not covariant. Using the explicit
expressions of $D_{0,\pm }$ and the local transformations (\ref{28}), we get
the following,%
\begin{eqnarray}
D_{+} &=&D_{+}^{\prime }+2\left( D_{+}\ln \Lambda \right) D_{0}^{\prime } 
\notag \\
D_{-} &=&D_{-}^{\prime }+2\left( D_{-}\ln \Lambda \right) D_{0}^{\prime } \\
D_{0} &=&D_{0}^{\prime }+2\left( D_{0}\ln \Lambda \right) D_{0}^{\prime }, 
\notag
\end{eqnarray}%
where one recognizes $2D_{0}$ as the generator of $\mathbb{C}^{\ast }$
isometry. To restore $\mathbb{C}^{\ast }$ gauge covariance, one should
introduce the holomorphic gauge fields 
\begin{equation}
A_{0,\pm }=A_{0,\pm }\left( \sigma ,x,y,z,w\right) ,
\end{equation}%
in order to covariantize the derivatives (\ref{15},\ref{18}) which becomes
then,%
\begin{eqnarray}
\mathcal{D}_{\pm } &=&D_{\pm }-A_{\pm }D_{0},  \notag \\
2\mathcal{D}_{0} &=&2D_{0}-2A_{0}D_{0},  \label{di}
\end{eqnarray}%
Note that on $T^{\ast }\mathbb{P}^{1}$, we should have $\mathcal{D}%
_{0}=D_{0} $, that is $A_{0}=0$ while on conifold such constraint equation
on the charge has no place. The gauge transformations of $A_{\pm }$ and $%
A_{0}$ are obtained by requiring $\mathcal{D}_{\pm }G_{n}$ to transform
covariantly 
\begin{eqnarray}
\mathcal{D}_{\pm }G_{n} &=&\Lambda ^{n\pm 2}\mathcal{D}_{\pm }G_{n},  \notag
\\
\mathcal{D}_{0}G_{n} &=&\Lambda ^{n}\mathcal{D}_{0}G_{n},
\end{eqnarray}%
which imply in turn the following gauge transformation of the gauge fields 
\begin{eqnarray*}
A_{\pm }G_{n} &\rightarrow &\Lambda ^{\pm 2}\left[ A_{\pm }+n\mathcal{D}%
_{\pm }\left( \ln \Lambda \right) \right] \left( \Lambda ^{n}G_{n}\right) ,
\\
A_{0}G_{n} &\rightarrow &\left[ A_{0}+n\mathcal{D}_{0}\left( \ln \Lambda
\right) \right] \Lambda ^{n}G_{n}.
\end{eqnarray*}%
They may be directly obtained from eqs(\ref{di}) by solving the constraint
eqs $\mathcal{D}_{\pm }^{\prime }=\Lambda ^{\pm 2}\mathcal{D}_{\pm }$ and $%
\mathcal{D}_{0}^{\prime }=\mathcal{D}_{0}$. We find, 
\begin{eqnarray}
A_{\pm } &\rightarrow &A_{\pm }^{\prime }=\Lambda ^{\pm 2}\left[ A_{\pm
}+D_{\pm }\left( \ln \Lambda \right) \right] ,  \notag \\
A_{0} &\rightarrow &A_{0}^{\prime }=A_{0}+2D_{0}\left( \ln \Lambda \right) .
\end{eqnarray}%
For the particular case of a global transformation where $\Lambda $ is
restricted to $\lambda $; that is $D_{0,\pm }\Lambda =D_{0,\pm }\lambda =0$,
one recovers the usual covariance of $A_{0,\pm }$ as holomorphic global
field sections.

These results for the fiber $T^{\ast }\mathbb{P}^{1}$ extend obviously to
the $SL\left( 2\right) $ generators $T_{0,\pm }$ on the base and more
generally to $\nabla _{0,\pm }$. On conifold, the previous fiber gauge
transformations read as,%
\begin{eqnarray}
\nabla _{\pm } &=&\nabla _{\pm }^{\prime }+2\left( \nabla _{\pm }\ln \Lambda
\right) \nabla _{0}^{\prime } \\
\nabla _{0} &=&\nabla _{0}^{\prime }+2\left( \nabla _{0}\ln \Lambda \right)
\nabla _{0}^{\prime }.  \notag
\end{eqnarray}%
The corresponding gauge covariant derivatives are given by,%
\begin{eqnarray}
\mathbf{\nabla }_{\pm } &=&\nabla _{\pm }-A_{\pm }\nabla _{0},  \notag \\
2\mathbf{\nabla }_{0} &=&2\nabla _{0}-2A_{0}\nabla _{0},
\end{eqnarray}%
and the gauge transformations of the gauge fields on conifold are as
follows, 
\begin{eqnarray}
A_{\pm } &\rightarrow &A_{\pm }^{\prime }=\Lambda ^{\pm 2}\left[ A_{\pm
}+\nabla _{\pm }\left( \ln \Lambda \right) \right] ,  \notag \\
A_{0} &\rightarrow &A_{0}^{\prime }=A_{0}+2\nabla _{0}\left( \ln \Lambda
\right) .
\end{eqnarray}%
With these tools at hand, we turn now to derive the correspondence between
conifold geometry and non commutative topological holomorphic $SL\left(
2\right) $ gauge theory.

\subsection{Deriving the topological gauge constraint eqs}

\qquad So far we have considered two coordinate systems to deal with
conifold geometry, an old free complex coordinate frame $\left\{
x_{i},y_{j}\right\} $ and a projective one $\left\{ \sigma ,x,y,z,w\right\} $%
. In the old coordinate system, conifold is seen as hypersurface $H_{0}$
embedded in $\mathbb{C}^{4}$; its defining eq $\varepsilon
^{ij}x_{i}y_{j}=\mu $ is invariant under the general coordinate change $%
x_{i}^{\prime }=y_{k}\Upsilon _{i}^{k}$, $y_{j}=\left( \Upsilon ^{-1}\right)
_{j}^{l}x_{l}$. Moreover, the generators of the global $SL\left( 2\right) $
isometry read as follows:%
\begin{eqnarray}
\nabla _{+} &=&\frac{1}{\sqrt{2}}\sum_{i}x_{i}\frac{\partial }{\partial y_{i}%
},  \notag \\
\nabla _{-} &=&\frac{1}{\sqrt{2}}\sum_{i}y_{i}\frac{\partial }{\partial x_{i}%
}, \\
\nabla _{0} &=&\frac{1}{2}\sum_{i}\left( x_{i}\frac{\partial }{\partial x_{i}%
}-y_{i}\frac{\partial }{\partial y_{i}}\right) .  \notag
\end{eqnarray}%
They satisfy the usual commutation relations%
\begin{equation}
\left[ \nabla _{+},\nabla _{-}\right] =2\nabla _{0},\qquad \left[ 2\nabla
_{0},\nabla _{\pm }\right] =\pm 2\nabla _{\pm }.  \label{nab}
\end{equation}%
Acting on the conifold hypersurface $x_{1}y_{2}-x_{2}y_{1}=\mu $ by these
operators, one discovers, as expected, the following relations,%
\begin{equation}
\left[ \nabla _{+},H\right] =\left[ \nabla _{-},H\right] =\left[ \nabla
_{0},H\right] =0.
\end{equation}%
Functions $\mathcal{F}_{R}\left( T^{\ast }S^{3}\right) $ on conifold
hypersurface $H_{0}$ are functions of the complex coordinates $x_{i}$ and $%
y_{j}$ with the restriction $\varepsilon ^{ij}x_{i}y_{j}=\mu $ and
transforming in representations $R$ of $SL\left( 2\right) $ isometry group.
Under the change $x_{i}^{\prime }=y_{k}\Upsilon _{i}^{k}$, $y_{j}=\left(
\Upsilon ^{-1}\right) _{j}^{l}x_{il}$, these functions $\mathcal{F}%
_{R}\left( T^{\ast }S^{3}\right) $ transform covariantly. For the
derivatives $\nabla _{0,\pm }\mathcal{F}_{R}$ to transform covariantly, one
has to introduce gauge connexions,%
\begin{equation}
\mathbf{\nabla }_{q}=\nabla _{q}-A_{q},\qquad q=0,\pm 1,
\end{equation}%
where the gauge fields $A_{q}=\sum_{p=0,\pm 1}B_{q}^{p}\nabla _{p}$ are
vector fields valued in $sl\left( 2\right) $ algebra. The $A_{0,\pm }$ gauge
fields are not all of them independent; their relations are obtained by
requiring that gauge covariant derivatives $\mathbf{\nabla }_{0,\pm }$
satisfy as well an $sl\left( 2\right) $ algebra, 
\begin{eqnarray}
\left[ \mathbf{\nabla }_{+},\mathbf{\nabla }_{-}\right] &=&2\mathbf{\nabla }%
_{0},  \notag \\
\left[ 2\mathbf{\nabla }_{0},\mathbf{\nabla }_{+}\right] &=&2\mathbf{\nabla }%
_{+},  \label{ch} \\
\left[ 2\mathbf{\nabla }_{0},\mathbf{\nabla }_{-}\right] &=&-2\mathbf{\nabla 
}_{-}.  \notag
\end{eqnarray}%
Notice that we have three gauge field components satisfying three constraint
eqs. This property signs the topological feature of $sl\left( 2\right) $
gauge theory on conifold. This description is also valid for in the
projective coordinate system $\left\{ \sigma ,x,y,z,w\right\} $ where
conifold is seen as a projective complex three surface embedded in $%
WP^{4}\left( {\small -1,1,-1,1,-1}\right) $. In this case, the previous
generators $\mathbf{\nabla }_{0,\pm }$ of the global $SL\left( 2\right) $
split as; 
\begin{equation}
\mathbf{\nabla }_{\pm }=\mathcal{D}_{\pm }-\mathcal{T}_{\pm },\qquad \mathbf{%
\nabla }_{0}=2\mathcal{D}_{0}-\mathcal{T}_{0},
\end{equation}
$\mathcal{T}_{0,\pm }$ for the base $\mathbb{C}^{\ast }$ and $\mathcal{D}%
_{0,\pm }$ for fiber $T^{\ast }\mathbb{P}^{1}$. Notice that that the
relations (\ref{ch}) may be put into a condensed form by help of the
completely antisymmetric invariant three dimensional tensor $\varepsilon
_{pqr}$ with the usual cyclic property $\varepsilon _{+-0}=\varepsilon
_{-0+}=\varepsilon _{0+-}=1$, one can put above eqs as follows, 
\begin{equation}
\left[ \mathbf{\nabla }_{p},\mathbf{\nabla }_{q}\right] =\varepsilon _{pqr}%
\mathbf{\nabla }_{-r},  \label{ab}
\end{equation}%
where we have renamed $2\mathbf{\nabla }_{0}$ and $\mathbf{\nabla }_{\pm }$
as $\mathbf{\nabla }_{0}^{\prime }$ and $\sqrt{\frac{1}{2}}\mathbf{\nabla }%
_{\pm }^{\prime }$ respectively; then dropped out the upper index prime. By
using the inverse tensor $\varepsilon ^{rpq}$ with the properties $%
\varepsilon ^{rpq}\varepsilon _{rqp}=2\delta _{s}^{r}$ and $\varepsilon
^{rpq}\varepsilon _{rqp}=6$, we can rewrite previous relation as follows,%
\begin{equation}
\varepsilon ^{rpq}\left[ \mathbf{\nabla }_{p},\mathbf{\nabla }_{q}\right] =-%
\frac{1}{2}\mathbf{\nabla }_{-r}
\end{equation}%
By substituting $\mathbf{\nabla }_{\pm }=\nabla _{\pm }-A_{\pm }\nabla _{0}$
and $\mathbf{\nabla }_{0}=\nabla _{0}-A_{0}\nabla _{0}$ back into the
constraint eqs(\ref{ab}), we get the following,%
\begin{eqnarray}
\left( \nabla _{+}A_{-}-\nabla _{-}A_{+}\right) +\left( A_{+}\nabla
_{0}A_{-}-A_{-}\nabla _{0}A_{+}\right) &=&A_{0},  \notag \\
\left( \nabla _{-}A_{0}-\nabla _{0}A_{-}\right) +\left( A_{-}\nabla
_{0}A_{0}-A_{0}\nabla _{0}A_{-}\right) &=&A_{-},  \label{na} \\
\left( \nabla _{0}A_{+}-\nabla _{+}A_{0}\right) +\left( A_{0}\nabla
_{0}A_{+}-A_{+}\nabla _{0}A_{0}\right) &=&A_{+},  \notag
\end{eqnarray}%
which read by help of gauge covariant derivatives $\mathbf{\nabla }_{p}$ in
a condensed form as,%
\begin{equation}
\varepsilon _{p}^{rpq}\mathbf{\nabla }A_{q}+A_{-r}=0.  \label{em}
\end{equation}%
Observe in passing that these constraint eqs are gauge invariant. Indeed
under gauge transformations $\delta A_{q}=$ $\mathbf{\nabla }_{q}\left( \ln
\Lambda \right) $, the quantity $\varepsilon ^{rpq}\left( \mathbf{\nabla }%
_{p}A_{q}\right) $ varies as $\varepsilon ^{rpq}\left[ \mathbf{\nabla }_{p}%
\mathbf{\nabla }_{q}\left( \ln \Lambda \right) \right] $ which by help of (%
\ref{ch}) we get $\frac{1}{2}\varepsilon ^{rpq}\varepsilon _{pq-s}\mathbf{%
\nabla }_{s}=-\delta A_{-r}$. Note finally that at first sight these
constraint eqs (\ref{na}) and analogs look a little bit unusual. While we
are dealing with an abelian gauge theory, our constraint eqs have generated
non linear terms. This is not a contradiction; it is just a signal of the
non commutative behavior of the topological $\mathbb{C}^{\ast }$ gauge
theory.

\subsection{The holomorphic topological action}

\qquad To get the complex holomorphic gauge field action $S_{T^{\ast }%
\mathbb{S}^{3}}=S_{T^{\ast }\mathbb{S}^{3}}\left[ A_{-},A_{+},A_{0}\right] $
describing the underlying NC topological holomorphic gauge theory, one
should think about the previous constraint relations (\ref{na}) as field
theoretic equation of motion following from the action principle%
\begin{equation}
\frac{\delta S_{T^{\ast }\mathbb{S}^{3}}}{\delta A_{r}}=0,\qquad r=0,+,-.
\end{equation}%
To solve this equation, it interesting to first express the action $%
S_{T^{\ast }\mathbb{S}^{3}}$ as the integral over a holomorphic integral
lagrangian like density $\mathcal{L}_{T^{\ast }\mathbb{S}^{3}}\left(
A\right) $ as,%
\begin{equation}
S_{T^{\ast }\mathbb{S}^{3}}=\frac{1}{\lambda }\int_{T^{\ast }\mathbb{S}%
^{3}}d^{3}v\mathcal{L}_{T^{\ast }\mathbb{S}^{3}}\left( A\right) ,
\end{equation}%
where $\lambda $ is the gauge coupling constant and where the three form $%
d^{3}v$\ stands for the conifold holomorphic volume form which, in the
projective coordinates $\left( \sigma ,x,y,z,w\right) $, splits as the
invariant volume 1-form of the $\mathbb{C}^{\ast }$ base times the invariant
volume 2-form of the fiber $T^{\ast }\mathbb{P}^{1}$. Projective and $%
SL\left( 2\right) $ invariances lead to the following local volume form,%
\begin{equation}
d^{3}v=\frac{d\sigma }{2i\pi \sigma }\times \left( dx\times dy-dz\times
dw\right)
\end{equation}%
Then equating $\frac{\partial \mathcal{L}_{T^{\ast }\mathbb{S}^{3}}}{\delta
A_{r}}$ with $\left( \varepsilon ^{rpq}\mathbf{\nabla }_{p}A_{q}+A_{-r}%
\right) $ eq(\ref{em}), which for convenience we rewrite it as%
\begin{equation}
\frac{\partial \mathcal{L}_{T^{\ast }\mathbb{S}^{3}}}{\delta A_{r}}%
=\varepsilon ^{rpq}\left[ \mathbf{\nabla }_{p}A_{q}-\frac{1}{2}\varepsilon
_{pqs}A_{-s}\right] ,
\end{equation}%
and integrating with respect to $A_{r}$, we obtain the following holomorphic
field Lagrangian density $\mathcal{L}\left( A\right) $, 
\begin{equation}
\mathcal{L}_{T^{\ast }\mathbb{S}^{3}}=\frac{1}{2}\varepsilon ^{rpq}\left[
A_{r}\mathbf{\nabla }_{P}A_{q}-\frac{1}{2}\varepsilon _{pqs}A_{r}A_{-s}%
\right] .
\end{equation}%
Gauge invariance follows naturally from covariance and the complete
antisymmetry of $\varepsilon ^{rpq}$ tensor. Note that while the first term
in bracket is the usual gauge term for abelian gauge theories, the second
one is typical for non commutative geometry.

\qquad On the real slice of the conifold, the initial $SL\left( 2,\mathbb{C}%
\right) $ symmetry reduces to $SU\left( 2,\mathbb{C}\right) $ and $\mathbb{C}%
^{\ast }$ to $U\left( 1\right) $ invariance, in particular we have the
following generators 
\begin{eqnarray}
\mathbf{\nabla }_{+}|_{S^{3}} &=&\frac{1}{\sqrt{2}}\left( x\frac{\partial }{%
\partial \overline{z}}-z\frac{\partial }{\partial \overline{x}}\right) , 
\notag \\
\mathbf{\nabla }_{-}|_{S^{3}} &=&\frac{1}{\sqrt{2}}\left( \overline{z}\frac{%
\partial }{\partial x}-\overline{x}\frac{\partial }{\partial z}\right) , \\
\mathbf{\nabla }_{0}|_{S^{3}} &=&\frac{1}{2}\left( x\frac{\partial }{%
\partial x}+z\frac{\partial }{\partial z}\right) -\left( \overline{x}\frac{%
\partial }{\partial \overline{x}}+\overline{z}\frac{\partial }{\partial 
\overline{z}}\right) .  \notag
\end{eqnarray}%
Similarly, the previous complex holomorphic gauge fields $A_{\pm }$ and $%
A_{0}$ reduce to $C_{\pm }$ and $C_{0}$; and obey the following reality
conditions,%
\begin{eqnarray}
\left( C_{\pm }\right) ^{\dagger } &=&C_{\mp },\qquad \left( C_{0}\right)
^{\dagger }=C_{0}\text{,}  \notag \\
\mathbf{\nabla }_{0,\pm } &=&\nabla _{0,\pm }-C_{0,\pm },
\end{eqnarray}%
while the constraint eqs read in the same manner as before,%
\begin{eqnarray}
\left[ \mathbf{\nabla }_{+},\mathbf{\nabla }_{-}\right] &=&\mathbf{\nabla }%
_{0},  \notag \\
\left[ \mathbf{\nabla }_{0},\mathbf{\nabla }_{+}\right] &=&\mathbf{\nabla }%
_{+}, \\
\left[ \mathbf{\nabla }_{0},\mathbf{\nabla }_{-}\right] &=&-\mathbf{\nabla }%
_{-}.  \notag
\end{eqnarray}%
It looks like $sl\left( 2\right) $ relations, except that now we have
moreover the reality conditions,%
\begin{equation}
\left( \mathbf{\nabla }_{\pm }\right) ^{\dagger }=\mathbf{\nabla }_{\mp
},\qquad \left( \mathbf{\nabla }_{0}\right) ^{\dagger }=\mathbf{\nabla }_{0}.
\end{equation}%
Following the same steps, we end with the non commutative topological
Chern-Simons gauge theory%
\begin{equation}
S_{\mathbb{S}^{3}}\left[ C\right] =\frac{1}{\lambda }\int_{\mathbb{S}^{3}}%
\frac{1}{2}\varepsilon ^{rpq}\left[ C_{r}\mathbf{\nabla }_{p}C_{q}-\frac{1}{2%
}\varepsilon _{pqs}C_{r}C_{-s}\right] ,
\end{equation}%
Gauge invariance follows in a similar way as for the complex holomorphic
case.

\subsubsection{Restrictions to $T^{\ast }\mathbb{P}^{1}$ and $\mathbb{S}^{2}$%
}

\qquad The above results extend naturally to the $T^{\ast }\mathbb{P}^{1}$
fiber sub-manifold. The point is that on this complex two dimensional non
compact projective surface of $W\mathbb{P}^{3}\left( {\small 1,-1,1,-1}%
\right) $, one can repeat quite same steps. In doing so, one should restrict
the previous covariant derivatives $\mathbf{\nabla }_{0,\pm }$ to the $%
T^{\ast }\mathbb{P}^{1}$ fiber; i.e 
\begin{equation}
\mathbf{\nabla }_{0,\pm }|_{T^{\ast }\mathbb{P}^{1}}=\mathcal{D}_{0,\pm },
\end{equation}
and moreover impose conservation of the $\mathbb{C}^{\ast }$ charge which
requires,%
\begin{equation}
\mathcal{D}_{0}=D_{0}\qquad \Leftrightarrow \qquad A_{0}=0.
\end{equation}%
As such there is no gauge component $A_{0}$ and no $\sigma $\ dependence.
Furthermore, generic functions $G_{n}=G_{n}\left( x,y,z,w\right) $ living on 
$T^{\ast }\mathbb{P}^{1}$ obey the following eigenvalue eq, 
\begin{equation}
2D_{0}G_{n}=nG_{n}.
\end{equation}%
In particular, we have for the holomorphic gauge fields $A_{\pm }$ on $%
T^{\ast }\mathbb{P}^{1}$, the following eigenvalue eqs, 
\begin{equation}
2D_{0}A_{\pm }=\pm 2A_{\pm }.
\end{equation}%
Unlike the identity $\mathcal{D}_{0}=D_{0}$, the two other gauge covariant
derivatives $\mathcal{D}_{\pm }$ keep their original form, 
\begin{equation}
\mathcal{D}_{\pm }=D_{\pm }-A_{\pm }D_{0}.
\end{equation}%
but with the gauge constraint eqs restricted to the fiber $T^{\ast }\mathbb{P%
}^{1},$%
\begin{eqnarray}
\left[ \mathcal{D}_{+},\mathcal{D}_{-}\right] &=&D_{0},  \notag \\
\left[ D_{0},\mathcal{D}_{+}\right] &=&\mathcal{D}_{+} \\
\left[ D_{0},\mathcal{D}_{-}\right] &=&-\mathcal{D}_{-}  \notag
\end{eqnarray}%
As such the previous three constraint relations (\ref{na}) reduce to the
following one,%
\begin{equation}
\left( D_{+}A_{-}-D_{-}A_{+}\right) -2A_{+}A_{-}=\left( \mathcal{D}_{+}A_{-}-%
\mathcal{D}_{-}A_{+}\right) =0,
\end{equation}%
together with \ the obvious identities, $D_{0}A_{\pm }=\pm A_{\pm }$. By
thinking about this constraint eq as a gauge field equation of motion
following from minimizing a gauge invariant holomorphic field action $%
S_{T^{\ast }\mathbb{P}^{1}}=S_{T^{\ast }\mathbb{P}^{1}}\left[
A_{-},A_{+},\Lambda _{0}\right] $ with respect to some some Lagrange field
parameter $\Lambda _{0}$; that is%
\begin{equation}
\frac{\delta S_{T^{\ast }\mathbb{P}^{1}}\left[ A_{-},A_{+},\Lambda _{0}%
\right] }{\delta \Lambda _{0}}=0,
\end{equation}%
one finds after integration, 
\begin{equation}
S_{T^{\ast }\mathbb{P}^{1}}=\frac{1}{\lambda }\int_{T^{\ast }\mathbb{P}%
^{1}}d^{2}v\Lambda _{0}\left( \mathcal{D}_{+}A_{-}-\mathcal{D}%
_{-}A_{+}\right) ,
\end{equation}%
where $d^{2}v$\ is the holomorphic volume form on $T^{\ast }\mathbb{P}^{1}$.
Clearly, this NC holomorphic $U(1)$ gauge field action $S_{T^{\ast }\mathbb{P%
}^{1}}$ is invariant under the gauge symmetry 
\begin{equation}
A_{\pm }\rightarrow A_{\pm }+\mathcal{D}_{\pm }\left( \ln \Lambda \right) .
\end{equation}%
One way to see it is by computing $\delta S_{T^{\ast }\mathbb{P}^{1}}\sim
\int_{T^{\ast }\mathbb{P}^{1}}\Lambda _{0}\left[ \mathcal{D}_{+},\mathcal{D}%
_{-}\right] \left( \ln \Lambda \right) $, which vanishes identically due to
the identity $D_{0}\left( \ln \Lambda \right) =0$. The reduction down to $%
\mathbb{S}^{2}$\ follows directly by imposing the reality condition. We get%
\begin{equation}
S_{\mathbb{S}^{2}}=\frac{1}{\lambda _{CS}}\int_{\mathbb{S}^{2}}d^{2}v\Lambda
_{0}\left( \mathcal{D}_{+}C_{-}-\mathcal{D}_{-}C_{+}\right)
\end{equation}%
where we have no gauge component $C_{0}$ and where $d^{2}v$ stands for the
real volume 2-form of the two sphere.

\subsubsection{Reductions to $T^{\ast }\mathbb{S}^{1}$ and $\mathbb{S}^{1}$}

\qquad The above results may be also reduced down to the $T^{\ast }\mathbb{S}%
^{1}$ base sub-manifold. The gauge covariant derivatives on conifold $%
\mathbf{\nabla }_{0,\pm }$ when restricted to the base $T^{\ast }\mathbb{S}%
^{1}$ reduce to $\mathcal{T}_{0,\pm }$ with,%
\begin{equation}
\mathcal{T}_{\pm }=T_{\pm }\qquad \Leftrightarrow \qquad A_{\pm }=0,
\end{equation}%
and 
\begin{equation}
\mathcal{T}_{0}=T_{0}-A_{0}T_{0},
\end{equation}%
where the gauge field $A_{0}$ has now no dependence of fiber variables, i.e, 
\begin{equation}
A_{0}=A_{0}\left( \sigma \right) .
\end{equation}%
As such there is no gauge components $A_{\pm }$ and no $\left(
x,y,z,w\right) $\ dependence. The constraint eqs for gauge field on conifold
reduce to,%
\begin{equation}
T_{\pm }A_{0}=0,
\end{equation}%
and 
\begin{equation*}
A_{0}=0.
\end{equation*}%
By equating this last relation with the action principle$\ \frac{\delta 
\mathcal{S}_{T^{\ast }\mathbb{S}^{1}}}{\delta A_{0}}=0$, we get the
topological holomorphic action $S_{T^{\ast }\mathbb{S}^{1}}=S_{T^{\ast }%
\mathbb{S}^{1}}\left[ A_{0}\right] $ on the base sub-manifold $T^{\ast }%
\mathbb{S}^{1}$ which reads then as, 
\begin{equation}
\mathcal{S}_{T^{\ast }\mathbb{S}^{1}}=\frac{1}{\lambda }\int_{T^{\ast }%
\mathbb{S}^{1}}\frac{d\sigma }{\sigma }A_{0}^{2},
\end{equation}%
where there is no kinetic term in agreement with the topological nature ot
the theory. Clearly, this NC holomorphic $U(1)$ gauge field action $%
S_{T^{\ast }\mathbb{P}^{1}}$ is invariant under the gauge symmetry, 
\begin{equation}
A_{0}\rightarrow A_{0}+\sigma \frac{\partial \left( \ln \Lambda \right) }{%
\partial \sigma }.
\end{equation}%
On the the unit circle, $\sigma =e^{i\theta }$ with $0\leq \theta <2\pi $,
this topological action reduces to $\mathcal{S}_{\mathbb{S}^{1}}=\frac{1}{%
\lambda }\int_{\mathbb{S}^{1}}d\theta C_{0}^{2}$ which is just a constant.

\section{Conclusion and outlook}

\qquad In an attempt to look for new methods to approach NC topological
gauge theories involving conifold background, we have first developed a way
to study the link between conifold and non commutativity opening then a
window for dealing with these backgrounds by borrowing q-deformed quantum
mechanics methods. Then we have studied the explicit derivation of NC
topological $SL\left( 2\right) $ gauge theory on conifold and its
sub-varieties using $T^{\ast }\mathbb{S}^{3}$ isometries. To do so, we have
started by showing that conifold defining eq $xy-zw=\varepsilon
^{ij}x_{i}y_{j}=\mu $ may be viewed as just the non trivial relation of the
defining eqs 
\begin{equation}
\left[ z_{I},z_{J}\right] _{q}=B_{IJ},\qquad I,J=1,2,3,4,
\end{equation}%
of non commutative complex four dimension space; but with a very specific
magnetic field eqs(\ref{eqa},\ref{5}). In comparing our approach with known
results in literature, we have noted striking similarities with Susskind way
to approach quantum Hall systems and Ooguri-Vafa-Verlinde study of
Hartle-Hawking Wave-Function for Flux Compactifications. We have developed
the similarity with Susskind non commutative model for the Laughlin state of
fractional quantum Hall system, known also to be described by a NC Chern
Simons $U\left( 1\right) $ gauge theory in $\left( 2+1\right) $ dimensions.
Then we have made a step in relating this feature to the attractor mechanism
of $\cite{24}$ also known to have a strong link with non commutative
geometry in so called mini-superspace.

\qquad Moreover, using the group factorisation of conifold $SL\left(
2\right) $ isometry as $\mathbb{C}^{\ast }\times \left( SL\left( 2\right) /%
\mathbb{C}^{\ast }\right) $, we have developed the corresponding projective
hypersurface representation of conifold. This way of doing has the advantage
of being directly related to the moduli space of supersymmetric theories
whose simplest model, with a $U_{gauge}\left( 1\right) \times
SU_{global}\left( 2\right) $ symmetry, has been constructed in section 4.
The holomorphic superpotential $\mathcal{L}$ of this supersymmetric model
involves the set of chiral matter $\left( Q_{+\alpha },P_{-\beta },\Sigma
_{+},\Sigma _{-},\Phi ,\Upsilon \right) $, see also (\ref{tab}). The
holomorphic eqs of motion read as 
\begin{eqnarray}
Q_{+\alpha }P_{-\beta }\varepsilon ^{\alpha \beta }+\Sigma _{+}\Sigma _{-}
&=&\mu ,  \notag \\
\Sigma _{+}\Sigma _{-} &=&1,  \label{con}
\end{eqnarray}%
where, for simplicity, we have dropped out the coupling constants. The
apparition of the charged superfields $\Sigma _{+}$ and $\Sigma _{-}$\ is
one of the predictions of this construction.

\qquad\ Furthermore, thinking about conifold as a projective complex three
dimension hypersurface embedded in non compact $W\mathbb{P}^{5}\left( 
{\small 1,-1,1,-1,1,-1}\right) $, we have developed a method to get
topological gauge theory by focusing on the gauging the $\mathbb{C}^{\ast }$
projective isometry. We have also studied the reduction of $SL\left(
2\right) $ gauge model on conifold down to its complex two and one
dimensions sub-manifold $T^{\ast }\mathbb{P}^{1}$ and $T^{\ast }\mathbb{S}%
^{1}$; as well as their real slices. Details on these topological gauge
reductions are exposed in section 6.

\qquad In the end of this study, we would like to add a comment on higher
dimension extensions of these geometries. As far as non commutative
structure is concerned, conifold results may be extended to higher complex
geometries by using the method presented in this paper. A direct extension
concerns complex dimension $\left( 4n-1\right) $ symplectic manifolds $%
SO\left( 4n,\mathbb{C}\right) /SO\left( 4n-1,\mathbb{C}\right) $ describing
the hypersurface $\sum_{a=1}^{n}\left( x_{a}y_{a+n}-x_{a+n}y_{a}=\mu \right) 
$ embedded in $\mathbb{C}^{4n}$. This equation may be put in the form and
reads also as,%
\begin{equation}
\sum_{A,B=1}^{2n}\Omega ^{AB}\left( x_{A}y_{B}-x_{B}y_{A}\right) =\mu ,
\end{equation}%
where $\Omega ^{AB}$ is the usual antisymmetric tensor of symplectic
spinors. Following the method outlined in section 3, this relation may be
also put as $x_{A}y_{B}-x_{B}y_{A}\sim \mu \Omega _{AB}$ or equivalently%
\begin{equation}
x_{A}y_{B}-\mathcal{R}_{AB}^{CD}y_{C}x_{D}\sim \mu \Omega _{AB}
\end{equation}
describing the link between these geometries and q-deformed non commutative
geometry in complex $4n$ dimensions with a constant magnetic field\ $\mu
\Omega _{AB}$. The particular case $n=1$ corresponds to conifold geometry
discussed in this paper. The next geometry, namely%
\begin{equation}
\left( x_{1}y_{3}-x_{3}y_{1}\right) +\left( x_{2}y_{4}-x_{4}y_{2}\right)
=\mu ,  \label{nan}
\end{equation}%
has a complex dimension $7$, containing $\mathbb{S}^{7}$ as a real slice,
and a manifest $SP\left( 1,\mathbb{C}\right) $ isometry subgroup rotating
the symplectic spinor $x_{a}$ into $y_{b}$ and vice versa. This construction
could be relevant for M-theory compactifications and $G_{2}$ manifolds.

\qquad An other interesting extension deals with complex dimension $\left(
n^{2}-1\right) $ embedded in $\mathbb{C}^{n^{2}}$ parameterized by $n\times
n $ matrix $Z$. These geometries have a $SL\left( n,C\right) $ isometry
group and are described by the holomorphic order $n$ polynomial equation,%
\begin{equation}
\det Z=\mu .
\end{equation}%
This algebraic relation captures a kind of generalized q-deformed non
commutative structure \`{a} la Nambu bracket aiming the construction of
generalizations of the hamiltonian mechanics based on Poisson bracket and
usual commutator. Indeed, using the $n$ dimensional completely antisymmetric
tensor $\varepsilon _{i_{1}...i_{n}}$, we can bring the above equation to
the remarkable form,%
\begin{equation}
Z_{1[j_{1}}Z_{2j_{2}}...Z_{nj_{n}]}=-\frac{\mu }{N!}\varepsilon
_{i_{1}...i_{n}},
\end{equation}%
and all others vanish identically. To fix the ideas, one may consider the
complex eight algebraic geometry with $SL\left( 3,C\right) $ isometries.
This geometry has some particularizes. First it may be directly related to
non commutative extension of Nambu mechanics whose bracket is associated
with the determinant of real $3\times 3$ matrices $\cite{a}$. Second, it
could be also relevant for type II superstring and M-theory
compactificactions. Let us give some explicit details concerning this
specific example. Consider the $3\times 3$ complex holomorphic matrix
coordinate,%
\begin{equation}
Z_{ij}=\left( 
\begin{array}{ccc}
u_{1} & u_{2} & u_{3} \\ 
v_{1} & v_{2} & v_{3} \\ 
w_{1} & w_{2} & w_{3}%
\end{array}%
\right)
\end{equation}%
with $u_{i}=Z_{1i},$ $v_{i}=Z_{2i}$ and $w_{i}=Z_{3i}$. The natural
extension of conifold geometry is given by the complex eight dimension
hypersurface embedded in $\mathbb{C}^{9},$%
\begin{equation}
\det Z=\varepsilon ^{ijk}u_{i}v_{j}w_{k}=\mu ,  \label{nam}
\end{equation}%
where $\varepsilon ^{ijk}$ is the invariant three dimensional completely
antisymmetric tensor. Like in case of conifold, this geometry is also
singular for $\mu =0$. In this coordinate system, the global isometry group
is generated by,%
\begin{eqnarray}
D_{1} &=&u^{i}\frac{\partial }{\partial w^{i}},\qquad D_{-1}=w^{i}\frac{%
\partial }{\partial u^{i}},,\qquad h_{1}=\left[ D_{1},D_{-1}\right] ,  \notag
\\
D_{2} &=&v^{i}\frac{\partial }{\partial w^{i}},\qquad D_{-2}=w^{i}\frac{%
\partial }{\partial v^{i}},\qquad h_{2}=\left[ D_{2},D_{-2}\right] , \\
D_{3} &=&u^{i}\frac{\partial }{\partial v^{i}},\qquad D_{-3}=v^{i}\frac{%
\partial }{\partial v^{i}},\qquad h_{3}=\left[ D_{3},D_{-3}\right]
=h_{1}-h_{2},  \notag
\end{eqnarray}%
and the link to non commutative geometry \`{a} la Nambu is given by,%
\begin{eqnarray}
u_{[i}v_{j}w_{k]} &=&-\frac{\mu }{6}\varepsilon _{ijk},  \notag \\
u_{[i}u_{j}w_{k]} &=&u_{[i}v_{j}v_{k]}=u_{[i}w_{j}w_{k]}=0.
\end{eqnarray}%
Concerning lower dimension sub-manifolds of eq(\ref{nam}), \ there are many;
the natural one is obtained by reducing $SL\left( 3,\mathbb{C}\right) $ to $%
SL\left( 3,\mathbb{R}\right) $ or other sub-groups such as $SU\left( 3,%
\mathbb{C}\right) $ or also $T^{\ast }\mathbb{S}^{1}\times T^{\ast }\mathbb{S%
}^{3}$. An other class of sub-manifolds is given by the following complex
five dimension geometry,%
\begin{equation}
\sum_{i=1}^{3}u_{i}\mu ^{i}=\mu ,
\end{equation}%
with%
\begin{equation}
\left( v_{j}w_{k}-v_{k}w_{j}\right) =-\frac{1}{2}\mu ^{i}\varepsilon _{ijk},
\end{equation}%
where $\mu $ and $\mu ^{i}$ are four constant numbers and where one
recognizes conifolds block as sub-geometries.

Following the method we have developed in section 4, one may also build
sub-manifolds with projective symmetries by help of the fibrations 
\begin{equation}
SL\left( 3\right) =\mathbb{C}^{\ast }\times \left( SL\left( 3\right) /%
\mathbb{C}^{\ast }\right) ,\qquad SL\left( 3\right) =\mathbb{C}^{\ast
2}\times \left( SL\left( 3\right) /\mathbb{C}^{\ast 2}\right) .
\end{equation}%
Using the group factorisation $SL\left( 3\right) =\mathbb{C}^{\ast }\times
\left( SL\left( 3\right) /\mathbb{C}^{\ast }\right) $, one can introduce the
projective $\left( \sigma ;x_{i},y_{j},z_{k}\right) $ related to the old
ones $\left( u_{i},v_{j},w_{k}\right) $ as follows,%
\begin{equation}
x_{i}=\sigma ^{q}u_{i},\qquad y_{j}=\sigma ^{p}v_{i},\qquad z_{k}=\sigma
^{r}w_{i},,\qquad q+p+r=0.
\end{equation}%
with the condition $q+p+r=0$ and, to fix the ideas, can be chosen as $q=p=1,$
$r=-2.$ This background has a natural supersymmetric quiver QFT$_{4}$
realization extending directly the model we have given for conifold. For
instance, the first relation of the superfield eqs of motion (\ref{con})
extends directly as 
\begin{eqnarray}
X_{+i}Y_{+j}Z_{-2k}\varepsilon ^{ijk}+\Sigma _{+}\Sigma _{-} &=&\mu ,  \notag
\\
\Sigma _{+}\Sigma _{-} &=&1,
\end{eqnarray}%
with $\Sigma _{+}\Sigma _{-}=1$ associated with $T^{\ast }\mathbb{S}^{1}$\
and $X_{+i}Y_{+j}Z_{-2k}\varepsilon ^{ijk}=\mu $ with the coset $SL\left(
3\right) /\mathbb{C}^{\ast }$. Quite similar relations may written down for
the others. For the fibration $SL\left( 3\right) =\mathbb{C}^{\ast 2}\times
\left( SL\left( 3\right) /\mathbb{C}^{\ast 2}\right) $, one may also build
the projective hypersurface and the corresponding supersymmetric QFT$_{4}$
realization by following the same method. The projective coordinates $\left(
\sigma ,\tau ;x_{i},y_{j},z_{k}\right) $ are related to the old ones $\left(
u_{i},v_{j},w_{k}\right) $ as follows,%
\begin{equation}
x_{i}=\sigma ^{q_{1}}\tau ^{q_{2}}u_{i},\qquad y_{j}=\sigma ^{p_{1}}\tau
^{p_{2}}v_{i},\qquad z_{k}=\sigma ^{r_{1}}\tau ^{r_{2}}w_{i},
\end{equation}%
with the two projective conditions $q_{a}+p_{a}+r_{a}=0$. Under the scaling
symmetries $\sigma \rightarrow \lambda _{1}\sigma $ and $\tau \rightarrow
\lambda _{2}\tau $, the new coordinates transforms as, 
\begin{equation}
x_{i}\rightarrow \lambda _{1}{}^{q_{1}}\lambda _{2}^{q_{2}}x_{i},\qquad
y_{j}\rightarrow \lambda _{1}{}^{p_{1}}\lambda _{2}{}^{p_{2}}y_{i},\qquad
z_{k}\rightarrow \lambda _{1}{}^{r_{1}}\lambda _{2}{}^{r_{2}}z_{k}
\end{equation}%
To build the corresponding $U_{gauge}\left( 1\right) \times U_{gauge}\left(
1\right) \times SU_{global}\left( 3\right) $ supersymmetric quiver gauge
theory, one should specify the solution of the constraint eqs $%
q_{a}+p_{a}+r_{a}=0$ \ and follows the same line as in the $SL\left(
2\right) $ case developed previously. In the special case $%
q_{a}+p_{a}=r_{a}=0$, the superfield degrees of freedom extending (\ref{tab}%
) are summarized in the table below,

\begin{equation}
\begin{tabular}{|l|l|}
\hline
4D $\mathcal{N}=1$ Superfields & $U\left( 1\right) \times U\left( 1\right)
\times SU\left( 3\right) $ \\ \hline
$V_{a}=-\theta \sigma ^{\mu }\overline{\theta }A_{\mu }+...,$ \ $a=1,2$ & $%
\left( 0,0,1\right) $ \\ \hline
$\Phi _{a}=\mathrm{\phi }_{a}+\theta \mathrm{\psi }_{a}+\theta ^{2}\mathrm{F}%
_{a}$ & $\left( 0,0,1\right) $ \\ \hline
$X_{i}=x_{i}+\theta \chi _{i}+\theta ^{2}F_{i}$ & $\left( 1,-1,3\right) $ \\ 
\hline
$Y_{i}=y_{i}+\theta \chi _{i}^{\prime }+\theta ^{2}F_{i}^{\prime }$ & $%
\left( -1,1,3\right) $ \\ \hline
$Z_{i}=z_{i}+\theta \chi _{i}^{\prime \prime }+\theta ^{2}F_{i}^{\prime
\prime }$ & $\left( 0,0,3\right) $ \\ \hline
$\Sigma _{\pm 1}=\sigma _{\pm 1}+\theta \eta _{\pm 1}+\theta ^{2}L_{\pm 1}$
& $\left( \pm 1,0,1\right) $ \\ \hline
$\Sigma _{\pm 2}=\sigma _{\pm 2}+\theta \eta _{\pm 2}+\theta ^{2}L_{\pm 2}$
& $\left( 0,\pm 1,1\right) $ \\ \hline
$\Upsilon _{0a}=\gamma _{0a}+\theta \tau _{0a}+\theta ^{2}G_{0a}$ & $\left(
0,0,1\right) $ \\ \hline
\end{tabular}%
\end{equation}%
Other superfield configurations are also possible. We end this discussion by
noting that it would be interesting to explore further the sub-manifolds of
eqs(\ref{nan},\ref{nam}) and look if they could be related with $G_{2}$
manifolds.

\begin{acknowledgement}
\qquad This work has been initiated during my visit to ICTP. I would like to
thank the Senior Associate program of ICTP for kind hospitality and
generosity. This research work is supported by the program Protars III
D12/25, CNRST.
\end{acknowledgement}


\begin{thebibliography}{99}
\bibitem{1} Edward Witten, Ground Ring Of Two Dimensional String Theory,
Nucl.Phys. B373 (1992) 187-213, hep-th/9108004,\newline
Edward Witten, Barton Zwiebach, Algebraic Structures and Differential
Geometry in 2D String Theory, Nucl.Phys. B377 (1992) 55-112, hep-th/9201056.

\bibitem{2} Igor R. Klebanov, String Theory in Two Dimensions, lectures at
the 1991 ICTP Spring School; hep-th/9108019

\bibitem{3} Gregory Moore, M. Ronen Plesser, Sanjaye Ramgoolam, Exact
S-Matrix for 2D String Theory, Nucl.Phys. B377 (1992) 143-190, hep-th/9111035

\bibitem{4} R. Dijkgraaf, G. Moore, R. Plesser, The partition function of 2d
string theory, Nucl.Phys. B394 (1993) 356-382, hep-th/9208031

\bibitem{5} Debashis Ghoshal, Cumrun Vafa, c=1 String as the Topological
Theory of the Conifold, Nucl.Phys. B453 (1995) 121-128, hep-th/9506122

\bibitem{6} Mina Aganagic, Ken Intriligator, Cumrun Vafa, Nicholas P. Warner
The Glueball Superpotential, Adv.Theor.Math.Phys. 7 (2004) 1045-1101,
hep-th/0304271

\bibitem{7} R. Dijkgraaf, M.T. Grisaru, C.S. Lam, C. Vafa, D. Zanon,
Perturbative Computation of Glueball Superpotentials, Phys.Lett. B573 (2003)
138-146, hep-th/0211017

\bibitem{8} Freddy Cachazo, Cumrun Vafa, N=1 and N=2 Geometry from Fluxes,
hep-th/0206017

\bibitem{9} F. Cachazo, B. Fiol, K. Intriligator, S. Katz, C. Vafa, A
Geometric Unification of Dualities, Nucl.Phys. B628 (2002) 3-78,
hep-th/0110028

\bibitem{10} R. Ahl Laamara, M. Ait Ben Haddou, A Belhaj, L.B Drissi, E.H
Saidi, RG Cascades in Hyperbolic Quiver Gauge Theories, Nucl.Phys. B702
(2004) 163-188, hep-th/0405222

\bibitem{11} F. Cachazo, S. Katz, C. Vafa, Geometric Transitions and N=1
Quiver Theories, hep-th/0108120

\bibitem{12} Mina Aganagic, Andrew Neitzke, Cumrun Vafa, BPS Microstates and
the Open Topological String Wave Function, hep-th/0504054

\bibitem{13} Sergei Gukov, Albert Schwarz, Cumrun Vafa, Khovanov-Rozansky
Homology and Topological Strings, hep-th/0412243

\bibitem{14} Marcos Marino, Chern-Simons Theory and Topological Strings,
Rev.Mod.Phys. 77 (2005) 675-720, hep-th/0406005

\bibitem{15} Harald Ita, Harald Nieder, Yaron Oz, Tadakatsu Sakai,
Topological B-Model, Matrix Models, \$\TEXTsymbol{\backslash}hat\{c\}=1\$
Strings and Quiver Gauge Theories, JHEP 0405 (2004) 058, hep-th/0403256

\bibitem{16} Robbert Dijkgraaf, Cumrun Vafa, Matrix Models, Topological
Strings, and Supersymmetric Gauge Theories, Nucl.Phys. B644 (2002) 3-20,
hep-th/0206255

\bibitem{17} Giulio Bonelli, Alessandro Tanzini, Maxim Zabzine, On
topological M-theory, hep-th/0509175

\bibitem{18} Nathan Seiberg, Edward Witten, String Theory and Noncommutative
Geometry, JHEP 9909 (1999) 032, hep-th/9908142

\bibitem{19} Mina Aganagic, Hirosi Ooguri, Natalia Saulina, Cumrun Vafa,
Black Holes, q-Deformed 2d Yang-Mills, and Non-perturbative Topological
Strings, Nucl.Phys. B715 (2005) 304-348, hep-th/0411280

\bibitem{20} A. Belhaj, M. Hssaini, E. L. Sahraoui, E. H. Saidi, Explicit
Derivation of Yang-Mills Self-Dual Solutions on non-Commutative Harmonic
Space, Class.Quant.Grav. 18 (2001) 2339-2358, hep-th/0007137

\bibitem{21} David Berenstein, Robert G. Leigh, Non-Commutative Calabi-Yau
Manifolds, Phys.Lett. B499 (2001) 207-214,hep-th/0009209,\newline
Adil Belhaj, El Hassan. Saidi, On Non Commutative Calabi-Yau Hypersurfaces ,
Phys.Lett. B523 (2001) 191-198, hep-th/0108143, NC Calabi-Yau Orbifolds in
Toric Varieties with Discrete, .Phys. A38 (2005) 721-748, hep-th/0210167

\bibitem{22} Mohamed Bennai, El Hassan Saidi, Toric Varieties with NC Toric
Actions: NC Type IIA Geometry, Nucl.Phys. B677 (2004) 587-613,
hep-th/0312200, NC Calabi-Yau Manifolds in Toric Varieties with NC Torus
fibration, Phys.Lett. B550 (2002) 108-116, hep-th/0210073

\bibitem{23} El Hassan Saidi, NC Geometry and Fractional Branes,
Class.Quant.Grav. 20 (2003) 4447-4472, hep-th/0311245

\bibitem{24} Hirosi Ooguri, Cumrun Vafa, Erik Verlinde, Hartle-Hawking
Wave-Function for Flux Compactifications, hep-th/0502211

\bibitem{25} Hirosi Ooguri, Andrew Strominger, Cumrun Vafa, Black Hole
Attractors and the Topological String, Phys.Rev. D70 (2004) 106007?
hep-th/0405146

\bibitem{26} L. Susskind, The Quantum Hall Fluid and Non-Commutative Chern
Simons Theory, hep-th/0101029,

\bibitem{27} Andrew Neitzke, Cumrun Vafa, Topological strings and their
physical applications, hep-th/0410178

\bibitem{28} Sergei Gukov, Kirill Saraikin, Cumrun Vafa, A Stringy Wave
Function for an $S^{3}$ Cosmology, hep-th/0505204

\bibitem{29} Conifold in harmonic space and correlations functions in
stringy $S^{3}$ cosmology, Lab/UFR-HEP 0512, GNPHE/05012, VACBT/05012

\bibitem{30} Malika Ait Benhaddou, El Hassan Saidi, Explicit Analysis of
Kahler Deformations in 4D N=1 Supersymmetric Quiver Theories, Physics
Letters B575(2003)100-110, hep-th/0307103

\bibitem{31} Edward Witten,Perturbative Gauge Theory As A String Theory In
Twistor Space, Commun.Math.Phys. 252 (2004) 189-258, hep-th/0312171

\bibitem{32} A.El Rhalami, E.M. Sahraoui, E.H.Saidi, NC Branes and
Hierarchies in Quantum Hall Fluids, JHEP 0205 (2002) 004, hep-th/0108096,

\bibitem{33} Aziz El Rhalami, El Hassan Saidi, NC Effective Gauge Model for
Multilayer FQH, JHEP 0210:039,2002, hep-th/0208144 States

\bibitem{34} R. Ahl Laamara, A. ElRhalami, E.H Saidi, Work in progress

\bibitem{35} Alexios P. Polychronakos, Quantum Hall states as matrix
Chern-Simons theory, JHEP 0104 (2001) 011, hep-th/0103013

\bibitem{36} Ahmed Jellal, El Hassan Saidi, Hendrik B. Geyer, A Matrix Model
for $\nu _{k_{1}k_{2}}=\frac{k_{1}+k_{2}}{k_{1}k_{2}}$ Fractional Quantum
Hall States, hep-th/0204248

\bibitem{37} James Gates Jr, Ahmed Jellal, EL Hassan Saidi, Michael
Schreiber, Supersymmetric Embedding of the Quantum Hall Matrix Model, JHEP
0411 (2004), hep-th/0410070 075

\bibitem{a} Azzouz Awane, Introduction \`{a} la g\'{e}ometrie
k-symplectique, Afr.J.Math.Phys.1:125-135,2004.

\bibitem{ads} R. Ahl Laamara, L.B Drissi, E.H Saidi, D-string fluid in
conifold, I.\textbf{\ }Topological gauge model,\textbf{\ } Lab/UFR-PHE 0516
\end{thebibliography}
\end{document}